\documentclass[twocolumn,aps,showpacs,amsmath,amsfonts,amssymb,floatfix,pre]{revtex4}
\usepackage{graphicx}
\draft
\begin{document}

\title{Synchronized clusters in coupled map
networks: Self-organized and driven phase synchronization} 
\author {Sarika Jalan\footnote{e-mail: sarika@prl.ernet.in} and
R. E. Amritkar\footnote{e-mail: amritkar@prl.ernet.in}}
\address{Physical Research Laboratory, Navrangpura, Ahmedabad 380 009, India.}
  
\begin{abstract}
We study the synchronization of coupled maps
on a variety of networks including regular one and two 
dimensional networks, scale free networks, small world networks, tree
networks, and random networks. 
The dynamics is governed by a local nonlinear 
map for each node of the network and interactions connecting different 
nodes via the links of the network. For small coupling strengths nodes
show turbulent behavior but form phase synchronized clusters as
coupling increases. We identify two
different ways of cluster formation, self-organized clusters 
which have mostly
intra-cluster couplings and driven clusters which have mostly 
inter-cluster couplings. 
The synchronized clusters may be of dominant self-organized 
type, dominant driven type or mixed type depending on the type of
network and the parameters of the dynamics. We also observe ideal
clusters of both self-organized and driven type. There are some nodes
of the floating type that show intermittent behaviour between getting
attached to some clusters and evolving independently. The residence
times of a floating node in a synchronized cluster show an
exponential distribution.
We define different states of the coupled dynamics by
considering the number and type of synchronized clusters. For 
the local dynamics
governed by the logistic map we study the phase
diagram in the plane of the coupling constant ($\epsilon$) and the
logistic map parameter ($\mu$). For large coupling strengths and
nonlinear coupling we find
that the scale free
networks and the Caley tree networks lead to better cluster formation than 
the other types of networks with the same average connectivity. 
For most of our
study we use the number of connections of the order of the number of nodes 
which allows us to distinguish between the two mechanisms of cluster
formation. As the number of connections increases the number of nodes 
forming clusters and the size of the clusters in general increase.
\end{abstract}


\maketitle
\section{Introduction} 
Several complex systems have underlying structures that are 
described by networks or graphs and the study of such networks is 
emerging as one of
the fastest growing subject in the physics world 
\cite{Strogatz,rev-Barabasi}.
One significant discovery in the field of complex networks is
the observation that a number of naturally occurring large and complex 
networks come under some universal classes and they can
be simulated with simple mathematical models, viz small-world 
networks \cite{Watts}, scale-free networks \cite{scalefree} etc. 
These models are based on simple physical considerations and have
attracted a lot of attention from physics 
community as they give simple algorithms to generate
graphs which resemble several actual networks found in many 
diverse systems such as the nervous systems \cite{koch}, 
social groups \cite{social},
world wide web \cite{www}, metabolic networks \cite{metabolic},
food webs \cite{food} and citation networks \cite{citation}.

Several networks in the real world consist of dynamical elements interacting
with each other. These networks have a large number
of degrees of freedom. In order to understand the
behaviour of these systems we study the synchronization
and cluster formation of these dynamical elements evolving
on different networks and connected via the links of the networks. 

Synchronization and cluster formation lead to rich spatio-temporal
patterns when opposing tendencies compete; the nonlinear dynamics 
of the maps which in the chaotic regime tends to separate the orbits 
of different elements, and the couplings that tend to synchronize them.
There are several studies on coupled maps/oscillators on regular
lattices as well as globally coupled networks.
Coupled map lattices with nearest neighbor
or short range interactions show interesting spatio-temporal patterns,
and intermittent behavior \cite{LCM1,LCM2}. Globally coupled
maps (GCM) where each node is connected with all other nodes, show
interesting synchronized behavior \cite{GCM1}. Formation
of clusters or coherent behaviour and then loss of coherence are 
described analytically as well as numerically at
different places with different points of views
\cite{GCM2,GCM3,GCM4,GCM5,GCM6,GCM-phasedia}.
Chaotic coupled map lattices show
beautiful phase ordering of nodes
\cite{HChate-phase}.
There are also some studies on coupled maps on different types of networks. 
Refs. \cite{Kaneko-2000,H-Chate,P-Gade} shed some 
light on the collective
behavior of coupled maps/oscillators with local and non-local connections.
Random networks with large number of connections also show
synchronized behavior for large coupling strengths
\cite{random-net1,random-net2,random-net3}. 
There are some studies on
synchronization of coupled maps on Cayley tree \cite{Tree}, 
small-world networks \cite{Small-1,Small-2,Small-3}
and hierarchal organization \cite{CHAOS-2002}.
Coupled map lattice with sine-circle map gives synchronization plateaus
\cite{CML-circle}. Analytical stability condition for synchronization
of coupled maps for different types of linear and non-linear
couplings are also discussed in several papers
\cite{CML-stability1,CML-stability2,CML-stability3}.
Synchronization and partial synchronization
of two coupled logistic maps are discussed at length in
Ref. \cite{bi-partite1}.
Apart from this there are other studies that explore
different properties of coupled maps
\cite{REA-gade,CML-REA,CML-other1,CML-other2,CML-other3}.

Coupled maps have been found to be useful in several practical
situations. These
include fluid dynamics \cite{ex-GCM}, nonstatistical behavior in
optical systems \cite{ex-optical}, convection
\cite{ex-Rayleigh,ex-convection}, stock markets
\cite{ex-stock}, ecological systems \cite{ex-eco}, logic gates
\cite{ex-CML}, solitons \cite{ex-soliton} and c-elegans
\cite{ex-celegans}. 

Here we study the detailed dynamics of coupled maps on different
networks and investigate the mechanism of clustering and
synchronization properties of such dynamically evolving networks.
We explore the evolution of individual nodes with time and 
study the role of different connections in forming
the clusters of synchronized nodes in such coupled map networks (CMNs).

Most of the earlier studies of synchronized cluster formation have focused on
networks with large number of connections ($\sim N^2$).
 In this paper, we consider networks
with number of connections
of the order of $N$. This small number of connections
allows us to study the mechanism of synchronized cluster formation and 
the role that different connections play in synchronizing different nodes.
We identify two phenomena, driven and self-organized
phase synchronization \cite{sarika-REA1}.
The connections or couplings in the self-organized phase synchronized
clusters are mostly of the intra-cluster type while those in the
driven phasen synchronized clusters are mostly of
the inter-cluster type.
As the number of connections increases more and more nodes
are involved in cluster formation and also the coupling strength
region where clusters are formed increases in size.
For large number of connections, typically of
the order of $N^2$ and for large coupling strengths, mostly one phase synchronized cluster spanning all
the nodes is observed.

Depending on the number and type of clusters we define different
states of synchronized behaviour. For the local dynamics governed by
the logistic map, we study the phase diagram in the $\mu-\epsilon$
plane, i.e. the plane defined by the logisting map parameter and the 
coupling constant.

The paper is organized as follows.
In section II, we give the model for our coupled map networks. We also
define phase synchronization and synchronized clusters as
well as discuss the mechanisms of cluster formation. In
section III, we present our numerical results for synchronization in
different networks
and illustrate the mechanism of cluster formation. This section
includes the study of the phase diagram, lyapunov exponent plots, behavior of
individual nodes, dependence on number of connections, and behaviour
for different types of networks. Some universal
features of synchronized cluster formation are discussed in section
IV. Section V considers circle map. Section
VI concludes the paper.

\section{Coupled maps and synchronized clusters}

\subsection{Model of a Coupled Map Network (CMN)}

 Consider a network of $N$ nodes and $N_c$ connections 
(or couplings)
between the nodes. Let each node of the
network be assigned a dynamical variable $x^i, i=1,2,\ldots,N$.
The evolution of the dynamical variables can be written as
\begin{equation}
x^{i}_{t + 1} = (1 - \epsilon) f( x^{i}_t ) + \frac{\epsilon}{k_i}
\sum_{j=1}^N C_{ij} g( x^{j}_t ),
\label{coupleddyn}
\end{equation}
where $x^{i}_t$ is the dynamical variable of the $i$-th 
node at
the $t$-th time step and $\epsilon$ is the coupling strength. The
topology of the network is introduced through the adjacency matrix $C$ with
elements $C_{ij}$
taking values $1$ or $0$ depending upon whether $i$ and $j$ are 
connected or not. $C$ is a symmetric matrix with diagonal 
elements zero. $k_i = \sum C_{ij}$ is the degree of node $i$. The
factors $(1-\epsilon)$ in the first term and $k_i$ in the second term
are introduced for normalization.
The function $f(x)$ defines the local nonlinear map and 
the function $g(x)$ defines the nature of coupling between the nodes. 
Here we present detailed results for the logistic
map,
\begin{equation}
f(x) = \mu x (1 - x)
\end{equation}
governing the local dynamics. We have also
considered some other maps for local dynamics. We have studied
different types of linear and non-linear coupling functions and
here discuss the results in detail for the following two types
of coupling functions.
\begin{mathletters}
\begin{eqnarray}
g(x) &=& x\\ 
g(x) &=& f(x).
\end{eqnarray}
\end{mathletters}
We refer to the first type of coupling function as linear and the
later as nonlinear.

\subsection{Phase synchronization and synchronized clusters}

Synchronization of coupled dynamical systems 
\cite{book-syn,phy-report,syn-coup} is indicated by the appearance of some
relations between the functionals of different dynamical variables due to
the interactions. The exact synchronization corresponds to
the situation where the dynamical variables for different nodes have identical
values. The phase synchronization corresponds the situation where the 
dynamical variables 
for different nodes have some definite relation between the phases 
\cite{phase1,phase2,phase3,phase4}. When the number 
of connections in the network
is small ($\sim N$) and when the local dynamics of the
nodes (i.e. function $f(x)$) is in the chaotic zone, only few
clusters with small number of nodes show exact synchronization.
However, clusters with larger number of nodes are obtained when we 
study phase synchronization. For our study we define the 
phase synchronization as follows \cite{syn}. 

Let $\nu_i$ and $\nu_j$ denote
the number of times the dynamical variables $x^i_t$ and $x^j_t$,
$t=t_0,t_0+1,2,\ldots,t_0+T-1$, for the nodes $i$ and $j$ show local 
minima during the time interval $T$ starting from some time
$t_0$. Here the local minimum of $x^i_t$ at time $t$ is defined by the 
conditions $x^i_{t} < x^i_{t-1}$ and $x^i_{t} < x^i_{t+1}$.
Let $\nu_{ij}$ denote the number of times these local minima match
with each other, i.e. occur at the same time. We define the phase
distance, $d_{ij}$,
between the nodes $i$ and $j$ by the following relation \cite{note1},
\begin{equation}
d_{ij} = 1-\frac{\nu_{ij}}{{\rm max}(\nu_i ,\nu_j)}.
\label{phase-dist}
\end{equation}
Clearly, $d_{ij}=d_{ji}$. Also, $d_{ij}=0$
when all minima of variables $x^i$ and $x^j$ match with each other
and $d_{ij}=1$ when none of the minima match.
In Appendix A, we show that the above definition of phase distance satisfies 
metric properties. 
We say that nodes $i$ and $j$ are phase synchronized if $d_{ij}=0$,
and a cluster of nodes is phase synchronized if all the pairs of nodes
belonging to that cluster are phase synchronized.

\subsection{Mechanism of cluster formation} 

Now we consider the relation between the synchronized
clusters which are formed by the dynamical evolution of nodes and
 the coupling between the nodes of
network which is a static property for a given network represented by
the adjacency matrix.
Clustering is obviously because of the coupling between
the nodes of the network and may be achieved in two different ways \cite{sarika-REA1}. \\
(i) The nodes of a cluster can be synchronized because of intra-cluster
couplings. We refer to this as the self-organized synchronization.\\
(ii) Alternately, the nodes of a cluster can be synchronized because
of inter-cluster couplings. Here nodes of one cluster are driven by those
of the others. We refer to this as the driven
synchronization. \\ 
We are able to identify ideal clusters of both the types,
as well as clusters of the mixed type where both ways of
synchronization contribute
to cluster formation.
 We will discuss several examples to
illustrate both types of clusters.

\subsection{States of synchronized dynamics}
We find that as network evolves, it splits into several
phase synchronized clusters.
The states of the coupled evolving system on the basis of number of clusters
can be classified as in Ref. \cite{different-region}\\
(a) Turbulent state (I): All nodes behave chaotically with no cluster
formation.\\
(b) Partially ordered state (III): Nodes form a few clusters with some
nodes not attached to any clusters. \\
(c) Ordered state (IV): Nodes form two or more clusters with
 no isolated nodes
at all. This ordered state can be further divided into 3 substates
based on the nature of nodes belonging to a cluster. These partitions are
chaotic ordered state, quasi-periodic ordered state, and periodic
ordered state.\\
(d) Coherent state (V): Nodes form a single synchronized cluster. \\
(e) Variable state (II): Nodes form different states, partially ordered 
or ordered states depending on initial conditions.

In addition to the above definition of states depending on the number
of clusters, we further divide the states having synchronized clusters 
into
subcategories depending on the type of clusters i.e. self-organized (S),
driven (D) or mixed type (M).

\section{Numerical results}
Now we present the numerical results of the coupled dynamics of variables 
associated with nodes on different types of networks. 
Starting from random initial conditions the 
dynamics of Eq.~(\ref{coupleddyn}),
after an initial transient, leads to interesting phase synchronized
behavior. The adjacency matrix $C$ depends on the type of network and
$C_{ij}=1$ if the corresponding nodes in the network are connected and 
zero otherwise.
First we will discuss our numerical results in detail for scale-free 
network and then we will discuss other networks.

\subsection{Coupled maps on scale-free network}

\begin{figure}
\centerline{
\includegraphics[width=9cm]{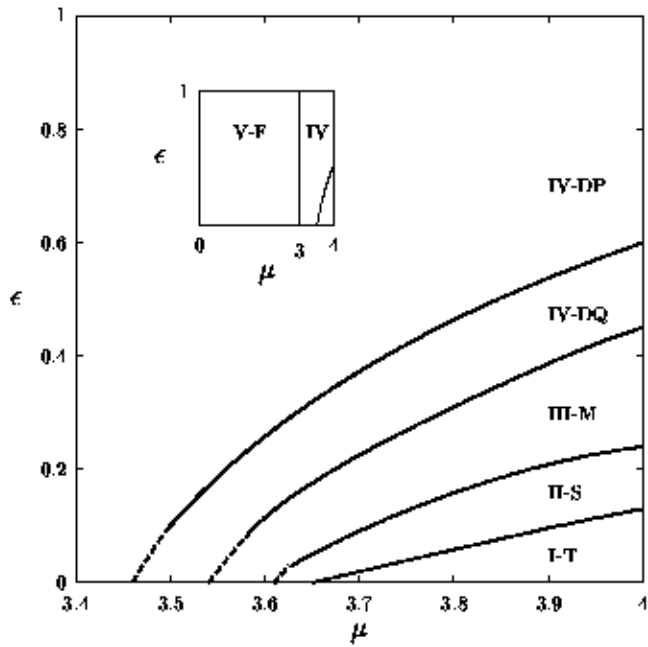}}
\caption{Phase diagram showing different regions in the two parameter
space of $\mu$ and
$\epsilon$ for scale free network for $f(x) = \mu x (1-x)$ and
$g(x)=x$. Different regions based on number
of clusters are I. Turbulent
region, II. region with varying behaviour, III. Partially ordered
region, IV. Ordered region, V. Coherent region. The
symbols T, S, M, D, P, Q and F respectively correspond to turbulent behaviour,
self-organized synchronization, mixed synchronization, driven
synchronization, periodic, quasiperiodic and fixed behaviour. Region
boundaries are determined based on
the asymptotic behaviour using several initial conditions,
number of clusters and isolated nodes, synchronization behaviour
and also the behaviour of
the largest Lyapunov exponent. The dashed
lines indicate uncertainties in
determining the boundaries.  Calculations
are for $N=50, m=1, T=100$. The inset shows the phase diagram for the entire
range of parameter $\mu$ i.e. from 0 to 4.}
\label{phase-scale-x}
\end{figure}

\subsubsection{Generation of Network}
The scale free network of $N$ nodes is generated by using the model of 
Barabasi et.al. \cite{model}.
Starting with a small number, $m_0$, of nodes, at each time
step a new node with $m \le m_0$ connections is added. The probability
$\pi(k_i)$ that a
connection starting from this new node is connected
to a node $i$ depends on the degree
$k_i$ of node $i$ (preferential attachment) and is given by
\[\pi(k_i) = \frac{(k_i + 1)}{\sum_j (k_j + 1)}.\] 
After $\tau$ time steps the
model leads to a network with $N = \tau + m_0$ nodes and $m \tau$
connections. This model leads to a scale free network, i.e. the
probability $P(k)$ that a node has a degree $k$ decays as a power law,
\[ P(k) \sim k^\lambda ,\]
where $\lambda$ is a constant and for the type of
probability law $\pi(k)$ that we have used $\lambda=3$. Other forms
for the probability $\pi(k)$ are possible which give different
values of $\lambda$. However, the results reported here do
not depend on the exact form of $\pi(k)$ except that it should lead to
a scale-free network.

\begin{figure}[ht]
\centerline{
\includegraphics[width=9cm]{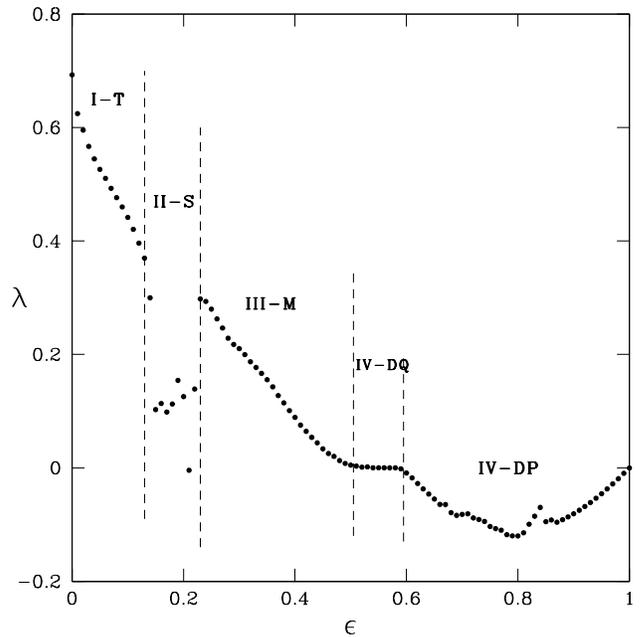}}
\caption{Largest Lyapunov exponent, $\lambda$, is plotted as a function of
$\epsilon$ for scale free network and $f(x) = 4 x (1-x)$ and
$g(x)=x$. Different regions are labeled
as in Fig.~\ref{phase-scale-x}.}
\label{lya-scale-x}
\end{figure}

\subsubsection{Linear coupling}
{\it Phase diagram}: First we start with the linear coupling, 
$g(x)=x$.
Fig.~\ref{phase-scale-x} shows the phase diagram in the two parameter space
defined by $\mu$ and $\epsilon$ for the scale-free network with
$m=m_0=1, N=50, T=100$. 
For $\mu<3$, we get a stable coherent region (region V-F) 
with all nodes having the
fixed point value. To understand the remaining phase diagram, consider
the line $\mu=4$. Fig.~\ref{lya-scale-x} shows the largest Lyapunov exponent
$\lambda$ as a function of the coupling strength $\epsilon$ for
$\mu=4$. We can identify four different regions
as $\epsilon$ increases from 0 to 1; namely the turbulent region,
the variable region (variable behaviour depending on $\epsilon$ 
and initial conditions), the partially ordered region 
and the ordered region
as shown by regions I to IV in Figs.~(1)
and~(2). The symbols T, S, M, DQ, DP and F correspond to turbulent,
self-organized, mixed, driven quasiperiodic, driven periodic and fixed 
point behaviour.
For small values of $\epsilon$, we observe the turbulent
behavior with all nodes evolving chaotically and there is no phase
synchronization (region I-T). There is a critical value of coupling
strength $\epsilon_c$ beyond which synchronized clusters can be
observed. This is a general property of all CMNs and
the exact value of $\epsilon_c$ depends on the type of network, the
type of coupling function and the parameter $\mu$. 

\begin{figure*}[ht]
\centerline{
\includegraphics[width=16cm,height=16cm]{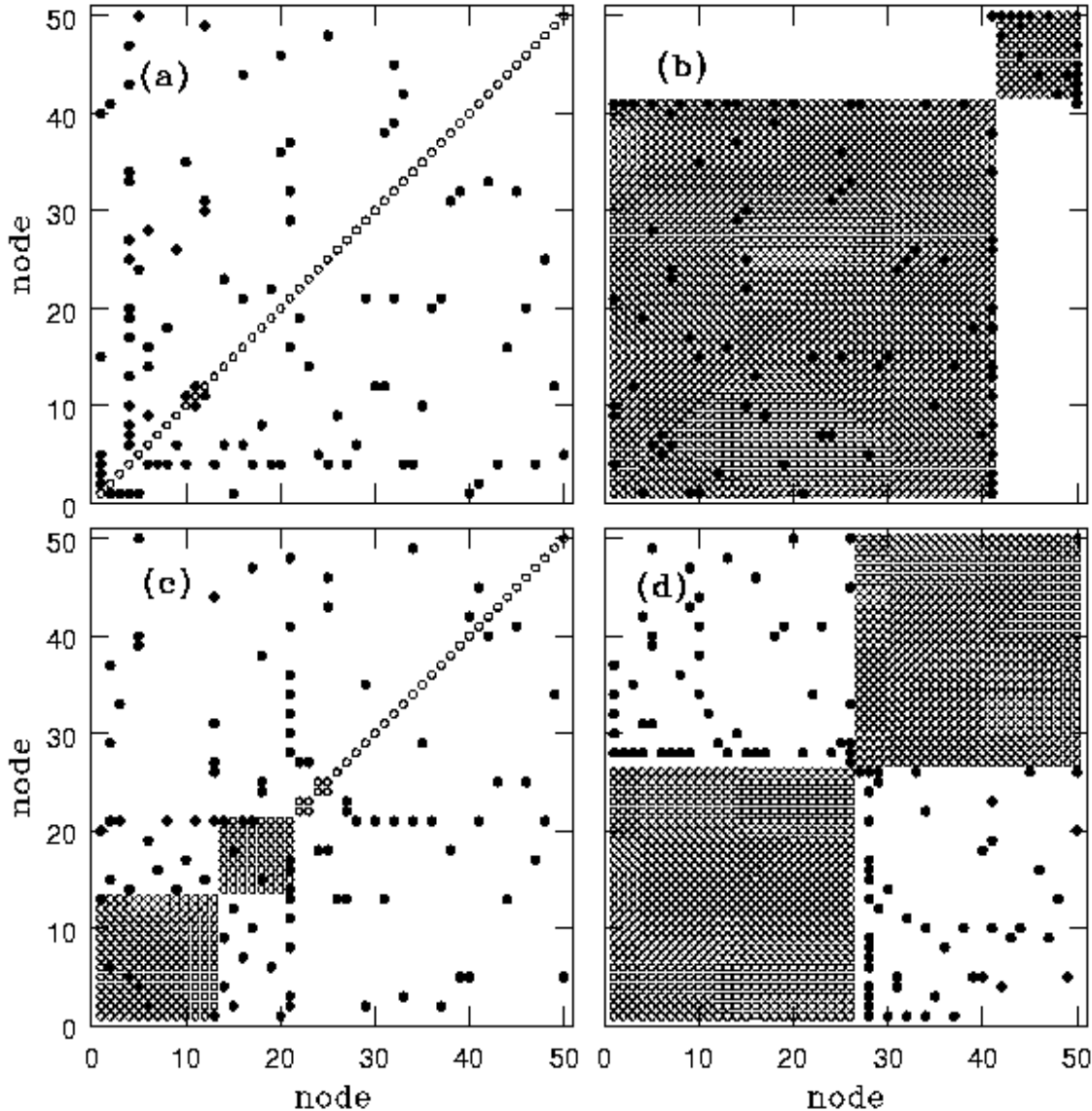}}
\caption{The figure shows several examples illustrating the
self-organized and driven phase synchronization. The examples are chosen
to demonstrate two different ways of obtaining synchronized clusters and the
variety of clusters that
are formed. All the figures show node verses node diagram for
$N=N_c=50$. After an initial transient (about 2000 iterates) pha$N=N_c=50$. Afte
r an initial transient (about 2000 iterates) phase
synchronized clusters are
studied for $T=100$. The logistic map parameter $\mu=4$ and coupstudied for $T=1
00$. The logistic map parameter $\mu=4$ and coupling
function $g(x)=x$. The solid
circles show that the two corresponding nodes are coupled and the open
circles show that the corresponding nodes are phase
synchronized. In each case the node numbers are reorganized so that
nodes belonging to the same cluster are
numbered consecutively and the clusters get displayed in decreasing
sizes. (a) Figure shows turbulent phase
for $\epsilon=0.10$.
(b) An ideal self-organized phase synchronization for $\epsilon=0.16$.
(c) Mixed behavior for
$\epsilon=0.32$. (d) A ideal driven phase synchronization
for $\epsilon=0.90$.
The scale free networks were generated with $N_0=1$ and $m=1$.}
\label{clus-scale-x}
\end{figure*}

As $\epsilon$ increases beyond $\epsilon_c$ we
get into a variable
region (region II-S) which shows a variety of phase synchronized behavior, 
namely ordered chaotic, ordered quasiperiodic, ordered periodic
 and partially 
orderedbehaviour depending on the initial conditions. The next region
(region III-M) shows partially ordered chaotic behavior. Here, 
the number of clusters as well as the number of nodes in the 
clusters depend on the initial conditions and also they change with time. 
There are several isolated nodes not belonging to any cluster. Many of these
nodes are of the floating type which keep on switching intermittently
between an independent evolution and a phase synchronized evolution
attached to some cluster. Last two regions (IV-DQ and IV-DP) are ordered 
quasiperiodic
and ordered periodic regions showing driven synchronization. In these
regions, the network always splits into 
two clusters. The two clusters are perfectly anti-phase synchronized
with each other, i.e. when the nodes belonging to one cluster show
minima those belonging to the other cluster show maxima.

We now investigate the nature of phase ordering in different regions
of the phase diagram. 
Fig.~\ref{clus-scale-x} shows node-node plots of the synchronized clusters
with any two nodes belonging to the same cluster shown as open
circles and the couplings between the nodes ($C_{ij}=1$) shown as solid
circles. For small coupling strength,
i.e. region I-T, nodes
show turbulent behaviour 
and no cluster is formed (Fig.~\ref{clus-scale-x}(a)). In region II-S the dominant
behaviour is of self-organized type. Fig.~\ref{clus-scale-x}(b) shows an ideal
self-organized synchronization with two clusters observed in the
middle of region II-S. Here, we observe that all the
couplings except one are of intra-cluster
type. Exactly opposite behavior is observed for the
regions IV-DQ and IV-DP. Fig. \ref{clus-scale-x}(d) shows an ideal 
driven synchronization obtained in
the middle of region IV-DP. Here, we find that all the couplings are 
of inter-cluster type with no intra-cluster couplings. This is clearly 
the phenomena of {\it driven synchronization} where the nodes of one cluster 
are driven into a phase synchronized state due to the couplings with 
the nodes of the other cluster. The phenomena of driven synchronization in 
this region is a very robust one
in the sense that it is obtained for almost all initial conditions, the
transient time is very small, the nodes belonging to the two clusters
are uniquely determined and we get a stable solution.
In region III-M we get clusters of mixed type (Fig.~\ref{clus-scale-x}(c)), 
here the inter-cluster connections and the intra-cluster connections are almost 
equal in numbers.

{\it Mechanism of cluster formation}:
We observe that for small values of $\epsilon$ the self organized
behavior dominates while for large $\epsilon$ driven behavior dominates.
As the coupling parameter $\epsilon$ increases from zero and we enter
the region II-S, we observe phase synchronized clusters of the self organized
type. Region III-M acts as a
crossover region from the self-organized to the driven behavior. Here, the
clusters are of the mixed type. The number of inter-cluster couplings
is approximately same as the number of intra-cluster couplings.
In this region there is a competition between the self-organized and driven 
behavior. This appears to be the reason for the formation of several
clusters and floating nodes as well as the sensitivity of these to the 
initial conditions. As $\epsilon$ increases, we get into region IV-DQ
where the driven
synchronization dominates and most of the connections between the nodes
are of the inter-cluster type with few intra-cluster connections. This driven
synchronization is further stabilized in region IV-DP with two
perfectly anti-phase synchronized driven clusters.

{\it Quantitative measure for self-organized and driven behaviour}:
To get a clear picture of self-organized and driven behaviour
we define two quantities $f_{\rm intra}$ and $f_{\rm inter}$ as
measures for the intra-cluster couplings and the inter-cluster couplings
as follows:
\begin{mathletters}
\begin{eqnarray}
f_{\rm intra} &=& \frac{N_{\rm intra}}{N_c}\\
f_{\rm inter} &=& \frac{N_{\rm inter}}{N_c}
\end{eqnarray}
\end{mathletters}
where $N_{\rm intra}$ and $N_{\rm inter}$ are the numbers of intra- and
inter-cluster couplings respectively. In $N_{\rm inter}$, couplings
between two isolated nodes are not included.

\begin{figure}[ht]
\centerline{
\includegraphics[width=9cm]{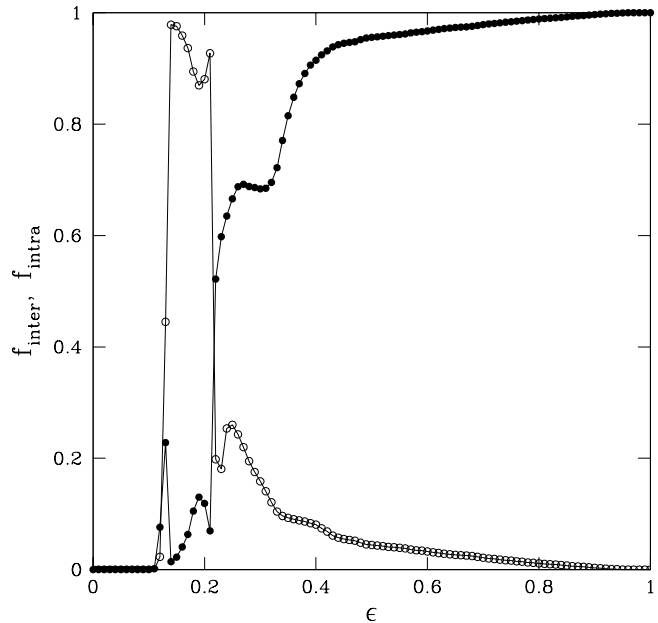}}
\caption{The fraction of intra-cluster and inter-cluster couplings,
$f_{inter}$ (solid circles) and $f_{intra}$ (open circle) are shown
as a function of the coupling
strength $\epsilon$ for the scale-free networks with $f(x)=4x(1-x)$
and $g(x)=x$.
The figure is obtained by averaging over 20
realizations of the network and 50 random initial conditions for
each realization.}
\label{inter-intra-scale-x}
\end{figure}

Fig.~\ref{inter-intra-scale-x} shows the plot of $f_{\rm intra}$ and
$f_{\rm inter}$ as a function of the coupling strength $\epsilon$.
The figure clearly shows that for small coupling
strength (region I-T) both $f_{\rm intra}$ and $f_{\rm inter}$ are zero
indicating that
there is no cluster formation at all, this is the turbulent region. As 
the
coupling strength increases ($\epsilon$ greater than some critical
value $\epsilon_c$) we get $f_{\rm intra} \sim 1$ at
$\epsilon \sim 0.2$ (region II-S). It shows that 
there are only intra-cluster couplings leading to
self-organized clusters. As coupling
strength increases further $f_{\rm intra}$ decreases and $f_{\rm inter}$ 
increases i.e. there is a crossover from self-organized to
driven behavior (regions III-M). As coupling strength 
enters  regions IV-DQ and IV-DP, we find that
$f_{\rm inter}$ is large which shows that in this region
most of the connections are of the inter-cluster type. In region IV-DP
we get  $f_{\rm inter}$ almost one
corresponding to an ideal driven synchronized behaviour.

\begin{figure*}
\centerline{\begin{tabular}{cc}
\includegraphics[width=9cm]{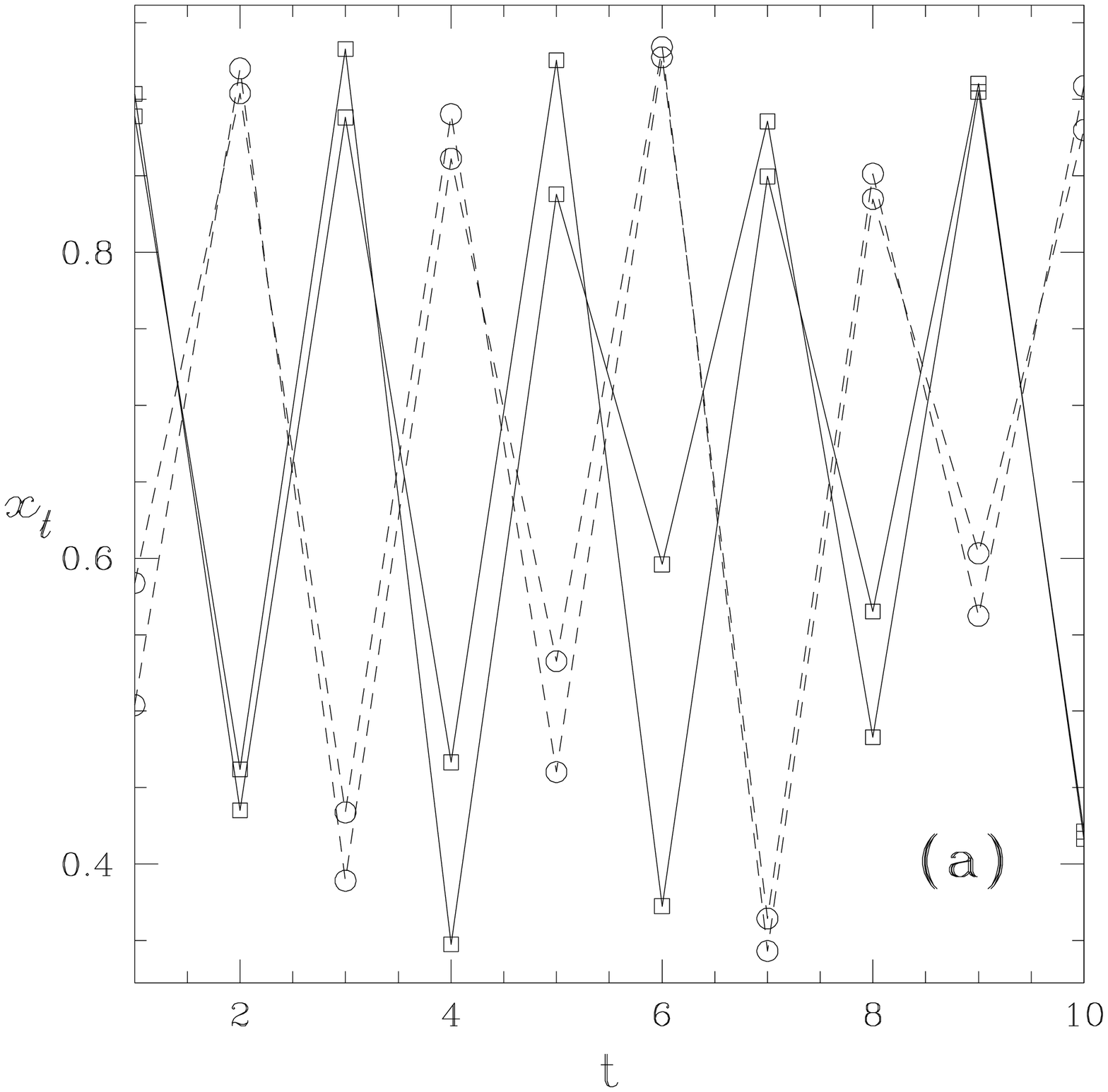} & \includegraphics[width=9cm]{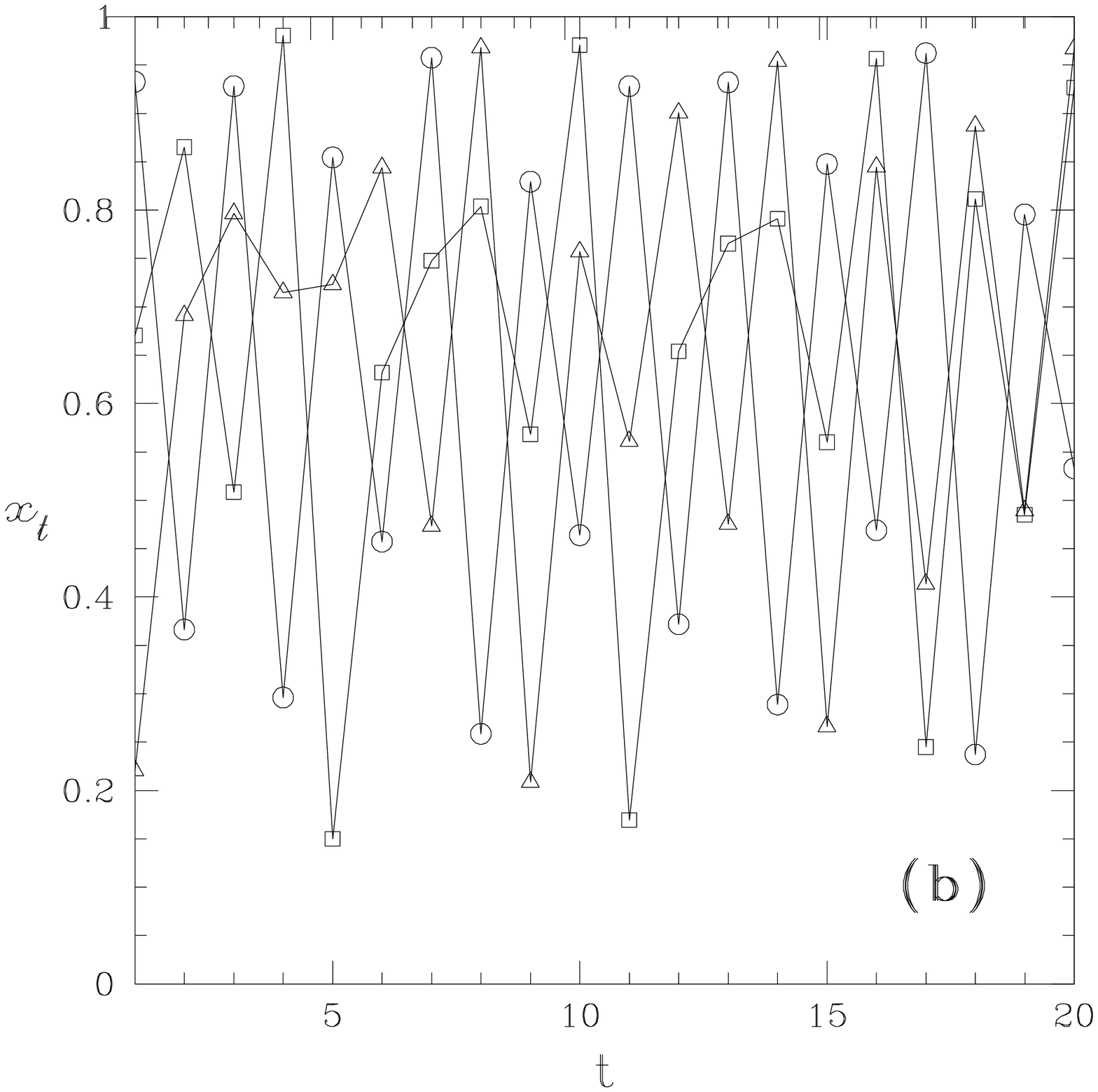}
\end{tabular} }
\caption{Figures show time evolution of nodes belonging to different clusters.
This figure is plotted for scale-free network with 50 nodes coupled with
$g(x) = x$. (a) A few nodes belonging to two phase synchronized clusters are
shown. Nodes denoted by
circles belong to one cluster and nodes denoted by squares to another
cluster. Here $\epsilon=0.15$. (b) Time series of
three nodes which are not phase synchronized with each other arethree nodes which are not phase synchronized with each other are shown with
three different symbols. Here, $\epsilon=0.35$.}
\label{tseries}
\end{figure*}

{\it Behaviour of individual nodes forming clusters}:
Figs.~\ref{tseries} (a) and (b) show plot of time evolution of some 
typical nodes. 
Fig.~\ref{tseries}(a) is for nodes in self-organized 
region ($\epsilon = 0.15$), where nodes belonging to the same cluster 
are marked 
with the same symbols. It is clearly seen that nodes with the same
symbols i.e. belonging to the same cluster are
phase synchronized and those belonging to different clusters are
completely anti-phase synchronized, i.e. when the nodes in one cluster are
showing minima, the nodes in other cluster are showing maxima.
(This behaviour is observed for driven behaviour where two clusters
are formed, i.e.
nodes belonging to different clusters are anti-phase synchronized with 
each other.) Fig.~\ref{tseries}(b) plots the time evolution of three nodes in the 
partially ordered region ($\epsilon = 0.35$). We see that these nodes are 
not phase synchronized with each other.

\begin{figure*}
\centerline{\begin{tabular}{ccc}
\includegraphics[width=9cm,height=9cm]{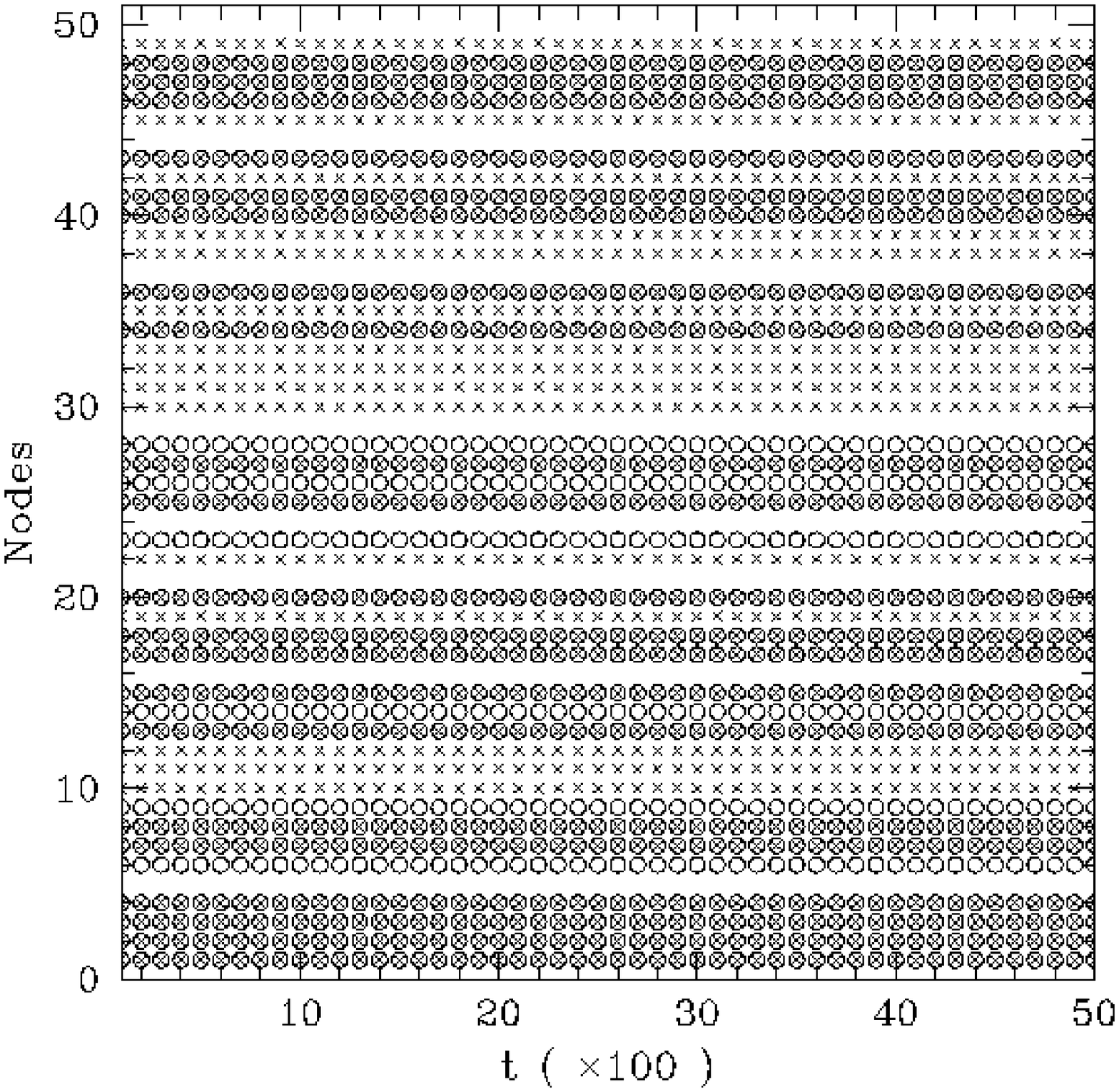} & \includegraphics[width=9cm,height=9cm]{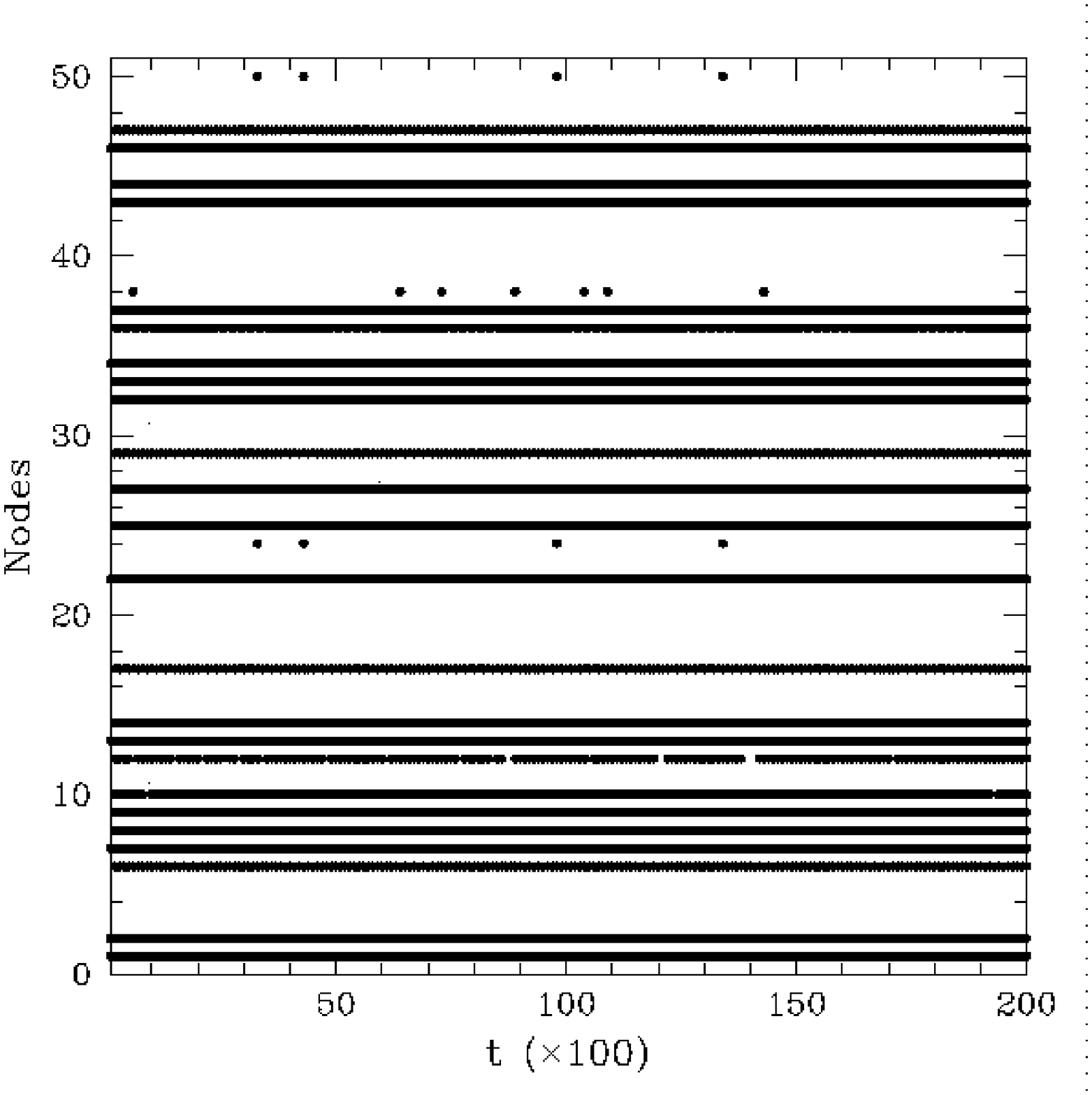}\\
\bf{(a)} & \bf{(b)}
\end{tabular} }
\caption{Figure shows the time evolution of nodes in a cluster for
scale-free network.
(a) shows two stationary clusters of self-organized type for
$\epsilon=0.19$ and $g(x)=x$. The two clusters are
for the same $\epsilon$ value but
for two different initial conditions. The nodes belonging to the two
clusters are denoted by open circles and crosses. Note that some
nodes are common to both the clusters while some are different .nodes are common to both the clusters while some are different .
This illustrates the nonuniqueness of nodes belonging
self-organized clusters depending on the initial conditions.
(b) shows a cluster
with some permanent nodes and some floating nodes. Here $\epsilon=0.4$
and $g(x)=x$.
Node number 12, 24, 38 and 50 are o$\epsilon=0.4$
and $g(x)=x$.
Node number 12, 24, 38 and 50 are of floating type.
They spend some time intermittently in a synchronized evolution with the
given cluster and the remaining time in either a synchronized evolution with
other clusters or in an independent evolution as an isolated node.}
\label{tseries-clus}
\end{figure*}

Now we explore different regions further to understand
time evolution of individual nodes attached to some specific cluster.
Fig.~\ref{tseries-clus} plots all the nodes belonging to a cluster as a
function of time, symbols indicate the time for which a given 
node belongs to a cluster.
First we consider region II-S. Fig.~\ref{tseries-clus}(a)
shows a set of nodes (crosses) belonging to a cluster 
in the mixed region for $\epsilon = 0.19$ and another set of nodes
(open circle) belonging to another cluster for the same $\epsilon$ 
but obtained with different initial conditions. 
For this $\epsilon$ value all the nodes form self-organized clusters
with no isolated nodes and this cluster formation is stable.
It is seen from Fig.~\ref{tseries-clus}(a) that all the nodes are 
permanent members of the cluster. Also comparing the members of two
clusters which 
are obtained from different initial conditions we see
that there are some common nodes while some are different. The
reason is that these are self-organized clusters and this organization 
is not unique (see subsection 3.A.4).
On the other hand, driven synchronization (region IV-DP) leads to a unique
cluster formation and does not depend on the initial conditions.

Next we look at $\epsilon=0.4$ (region III-M) where we get several
clusters with some isolated nodes. In Fig.~\ref{tseries-clus}(b),
nodes belonging to a cluster are plotted as a function of time. We observe 
that there are some nodes which are attached to this cluster, intermittently
leave the cluster, evolve independently or get attached
with some other cluster and after some time again come back to
the same cluster. These
nodes are of the floating type which keep on switching intermittently
between an independent evolution and a phase synchronized evolution
with some cluster. 
For example, node number 12 in Fig.~(tseries-clus)(b), which 
forms phase synchronized cluster with other nodes, 
in between leaves the cluster and evolve independently for some time.
Time it spends with the cluster is about 90\%. On the other hand
node number 24 evolves independently for almost 90\% of the time
and evolves in phase synchronization with the cluster for the 
rest of the time.

\begin{figure}
\centerline{
\includegraphics[width=9cm]{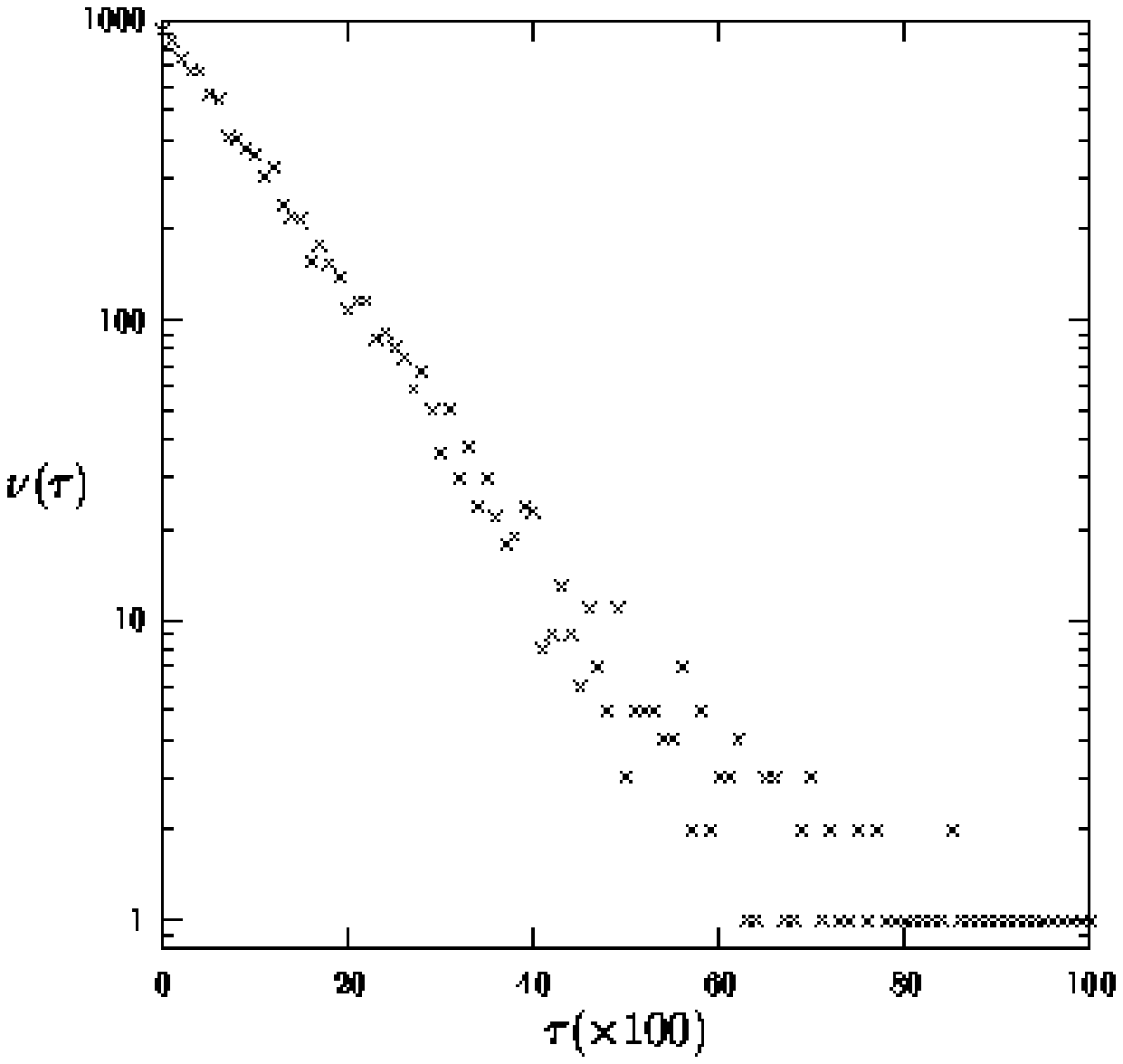}}
\caption{The figure plots the frequency of residence time $f(\tau)$
of a floating node in a cluster as a
function of the residence time $\tau$. The data is for node no 12
in Fig.~\ref{tseries-clus}(b). A good straight line fit on log-linear
plot shows exponential dependence.
}
\label{freq-floating}
\end{figure}

Let $\tau$ denote the residence time of a floating node in a cluster 
(i.e. the continuous time interval that the node is in a cluster).
Fig.~\ref{freq-floating} plots the frequency of residence time $f(\tau)$
of a floating node as a
function of the residence time $\tau$. 
A good straight line fit on log-linear
plot shows exponential dependence, $f(\tau) \sim \exp(-\tau/ \tau_r)$ where
$\tau_r$ is the typical residence time
for a given node.
 
\subsubsection{Nonlinear coupling}
\begin{figure}
\centerline{
\includegraphics[width=9cm]{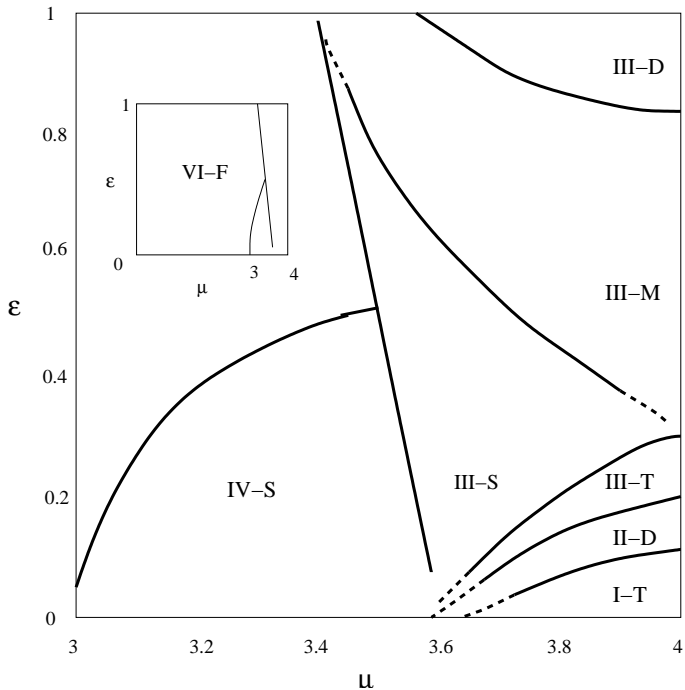}}
\caption{Phase diagram showing turbulent, phase synchronized and
coherent regions in the two parameter space of $\mu$ and
$\epsilon$ for scale free network for $f(x) = \mu x (1-x)$ and
$g(x)=f(x)$. The determination of region boundaries and their
classification and symbols are as explained in
Fig.~\ref{phase-scale-x}. Calculations
are for $N=50, m=1, T=100$. The inset shows the phase diagram for the
entire range of parameter $\mu$ i.e. from 0 to 4.}
\label{phase-scale-fx}
\end{figure}

\begin{figure}
\centerline{
\includegraphics[width=9cm]{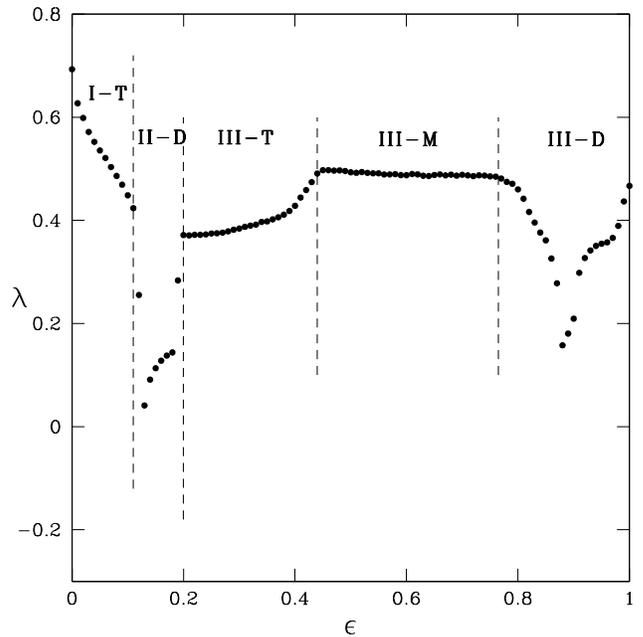}}
\caption{Largest Lyapunov exponent, $\lambda$, is plotted as a function of
$\epsilon$ for scale-free network and $f(x) = 4 x (1-x)$ and
$g(x)=f(x)$. Different regions are labeled as in Fig.~\ref{phase-scale-fx}.}
\label{lya-scale-fx}
\end{figure}

Now we discuss the results for the nonlinear coupling of
the type $g(x)=f(x)$. Phase space diagram in the $\mu - \epsilon$ plane
is plotted in Fig.~\ref{phase-scale-fx}.
Here we do not get clear and distinct regions as we get for
$g(x) = x$ form of coupling. Again the phase diagram is divided into
different regions I to V,
based on the number of clusters
as given in the beginning of this section.
For $\mu < 3.5$, we get coherent behaviour (regions V-P and VI-F).
To describe the remaining phase diagram first consider $\mu = 4$ line.
Figure \ref{lya-scale-fx} shows the largest Lyapunov exponent as a
function of the 
coupling strength $\epsilon$ for $\mu=4$.
For small coupling strengths no cluster is formed and we get the
turbulent region (I-T). As the coupling strength increases we get into
the variable region (II-D). In this region we get partially ordered and
ordered chaotic phase depending on
the initial conditions. 
In a small portion in the middle of region II-D,
all nodes form two ideal driven clusters. 
These two clusters are perfectly anti-phase 
synchronized with each other. Interestingly the dynamics still remains
chaotic. In region III-T, we get almost turbulent behaviour with very
few nodes forming synchronized clusters. Regions III-M and III-D are partially
ordered chaotic regions. In these regions some nodes form
clusters and several nodes are isolated or of the floating type.


\begin{figure}
\begin{center}
\includegraphics{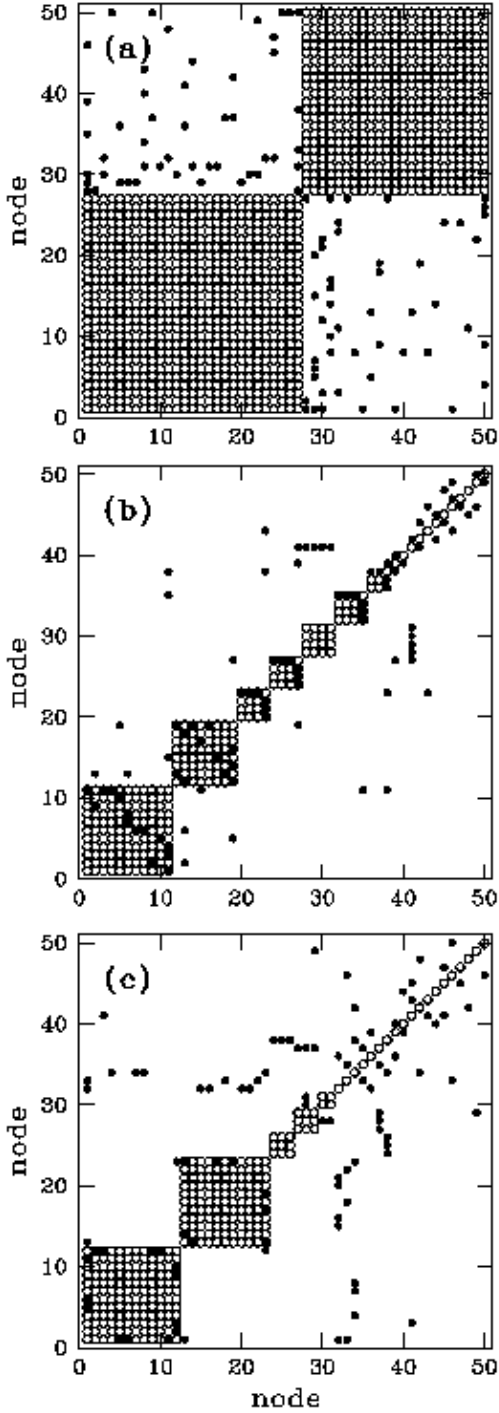}
\caption{The figure shows several examples illustrating the
phase synchronization for scale-free
network with coupling form $g(x)=f(x)$
using node verses node diagram for
$N=N_c=50$. After an initial transient (about 2000 iterates) phase
synchronized clusters are
studied for $T=100$. The logistic map parameter $\mu=4$. The solid
circles show that the two corresponding nodes are coupled and the open
circles show that the corresponding nodes are phase
synchronized. In each case the node numbers are reorganized so that
nodes belonging to the same cluster are
numbered consecutively and the clusters get displayed in decreasing
sizes. (a) Figure show an ideal driven phase
synchronization for $\epsilon=0.13$.
(b) Mixed behavior for
$\epsilon=0.71$. (c) A dominant driven behavior for $\epsilon=0.88$.
The scale free networks were generated with $N_0=1$ and $m=1$.}
\label{clus-scale-fx}
\end{center}
\end{figure}

We now investigate the nature of phase ordering in different regions
of the phase diagram. Figs.~\ref{clus-scale-fx} are node-node plots showing
different clusters and couplings (as in Fig.~\ref{clus-scale-x}) 
for different
$\epsilon$ values belonging to different regions.  
In Fig.~\ref{clus-scale-fx}(a) (region II-D) we observe that all the
couplings are of inter-cluster type with no intra-cluster coupling. This is
the phenomenon of driven synchronization. 
There are two clusters which are perfectly anti-phase synchronized. 
Fig.~\ref{clus-scale-fx}(b) is plotted for $\epsilon$ in region III-M
and it shows clusters of different types. 
The fifth cluster has only inter-cluster couplings (driven type) while 
the remaining clusters have dominant intra-cluster couplings 
(self-organized type).
There are several isolated nodes also. Fig.~\ref{clus-scale-fx}(c)
shows clusters where the driven behaviour dominates. 
It is interesting to note that for scale-free network and
for this type of nonlinear coupling largest Lyapunov exponent is always
positive (Fig.~\ref{lya-scale-fx}) i.e. the whole system remains chaotic but we 
get phase-synchronized behavior.

\begin{figure}
\begin{center}
\centerline{
\includegraphics[width=9cm]{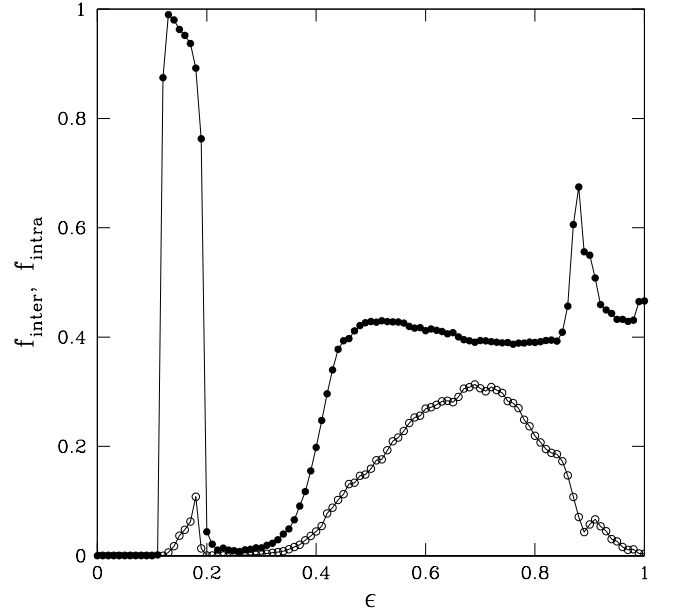}}
\caption{The fraction of intra-cluster and inter-cluster couplings,
$f_{inter}$ (solid circles) and $f_{intra}$ (open circle) are shown
as a function of the coupling
strength $\epsilon$ for the scale-free networks with $g(x)=f(x)$.
The figure is obtained by averaging over 20
realizations of the network and 50 random initial conditions for
each realization.}
\label{inter-intra-scale-fx}
\end{center}
\end{figure}

Fig.~\ref{inter-intra-scale-fx} shows the plot of $f_{\rm intra}$ and
$f_{\rm inter}$ as a function of the coupling strength $\epsilon$ for $\mu=4$.
For small coupling strength both quantities are
zero showing the turbulent region, and as the coupling strength increases
clusters are formed. $f_{\rm inter}$ is one at $\epsilon \approx 0.13$   
which shows that here all the nodes are forming clusters and the clusters 
are of the driven type having only inter-cluster
connections. As the coupling strength increases further, 
$f_{\rm inter}$ and $f_{\rm intra}$ become almost zero (region III-T)
and subsequently start increasing slowly (region III-M) but we see
that $f_{\rm inter}$ is
always greater than $f_{\rm intra}$ leading to dominant driven phase
synchronized clusters. For $\epsilon > 0.7$, $f_{\rm intra}$ starts
decreasing and for $\epsilon > 0.8$, the driven behaviour becomes more
prominent (region III-D and
Fig.~\ref{clus-scale-fx}(c)). For the regions III-M and III-D, we get
phase synchronized clusters
but the size of clusters as well as the number of nodes forming clusters
both are small.

\subsubsection{Network geometry and Cluster formation}

Geometrically, the organization of the scale-free network into 
connections of both
self-organized and driven types is always possible for $m=1$. For
$m=1$, our growth algorithm generates a tree type structure. A tree
can be broken into different clusters in two distinct ways. \\
(a) A tree can be broken into two or more disjoint clusters with only
intra-cluster couplings by breaking one
or more connections. Clearly, this splitting is not unique.
This behaviour is observed in region II-S of the phase-diagram in
Fig.~1 and
can be seen by comparing Fig. 3(b) of this paper (two ideal
self-organized cluster of sizes 41 and 9) and Fig. 1(a) of Ref.
\cite{sarika-REA1} (two ideal self-organized cluster of sizes 36 and
14) which are plotted for the same scale free network and $g(x)=x$
but for different $\epsilon$ values. \\
(b) A tree can 
also be divided into two clusters by putting connected nodes into different
clusters. This division is unique and leads to two clusters with only
inter-cluster couplings. This behaviour is observed in region IV-DP of
the phase-diagram in Fig.~1 and can be seen by comparing Fig. 3(d) of this paper and
Fig. 1(b) of Ref. \cite{sarika-REA1} which are again plotted for the same
scale free network but for diferrent $\epsilon$ values.  

\begin{figure*}
\centerline{
\includegraphics[width=18cm,height=18cm]{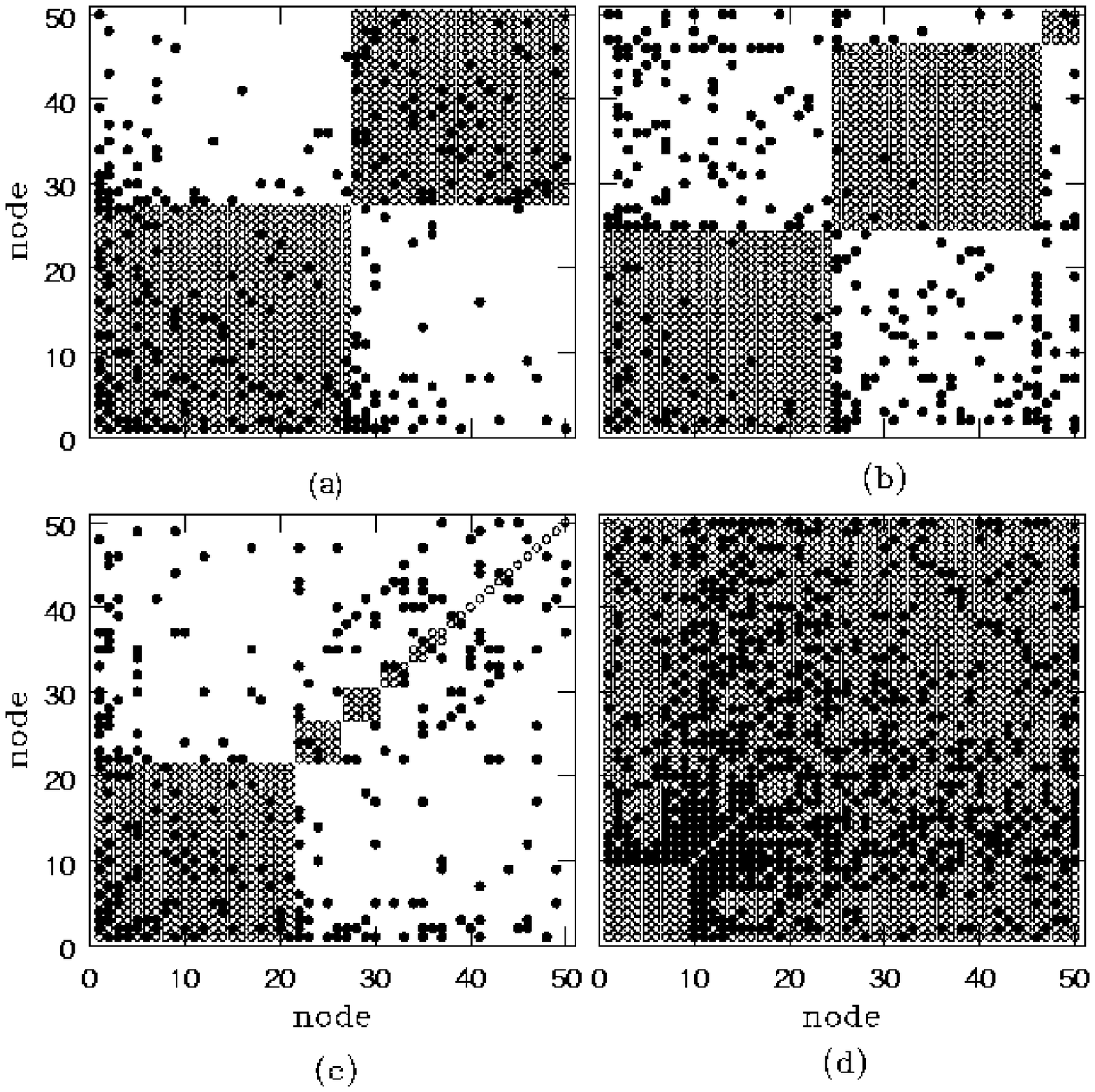}}
\caption{Figure illustrates the cluster formation for
the scale-free network as node vs node plot for $N=50$ as in
Fig.~\ref{clus-scale-x} but
with larger number of connections. (a) and (b) are plotted for
$g(x)=x$ and $m=3$ and respectively for $\epsilon = 0.19$
and $\epsilon = 0.78$. (c) and
(d) are plotted for $g(x) = f(x)$,  $\epsilon = 0.90$ and respectively
for $m=3$ and $m=10$.}
\label{clus-scale-m}
\end{figure*}

For $m >1$ and $g(x) = x$ the dynamics of Eq.~(\ref{coupleddyn}) leads 
to a similar phase diagram as in Fig.~(1) with region II-S dominated by
self-organized synchronization and regions IV-DQ and IV-DP dominated by 
driven
synchronization. Though perfect inter- and intra-cluster couplings between 
the nodes as displayed in Figs.~(3b) and ~(3d) are no longer
observed, clustering in the region II-S is such that most of the couplings are
of the intra-cluster type while for the regions IV-DQ and IV-DP they 
are of the inter-cluster type. As $m$ increases the
regions I and II are mostly
unaffected, but the region IV shrinks and the region III grows in size.
Fig.~\ref{clus-scale-m} shows different types of clusters in
node-node diagram for
different coupling strength. Fig.~\ref{clus-scale-m}(a) is plotted for
$m=3$ in the 
variable region ($\epsilon = 0.19$) with nodes forming two clusters.
It is clear that synchronization of nodes is mainly because
of intra-cluster connections but there are a few inter-cluster 
connections also.
Fig.~\ref{clus-scale-m}(b) is plotted for the ordered periodic region
at coupling strength $\epsilon = 0.78$, here the clusters are
mainly of the driven type but they have intra-cluster connections also.
In Figures \ref{clus-scale-m}(a) and~(b) the average degree of a node
is 6, and breaking the network into clusters
with only inter-cluster or intra-cluster couplings is not possible. As
the average degree of a node 
increases further self-organized behaviour starts dominating and for
most of the $\epsilon$ values nodes behave in 
phase-synchronized manner forming one big cluster.

\begin{figure*}
\centerline{
\includegraphics[width=12cm,height=12cm]{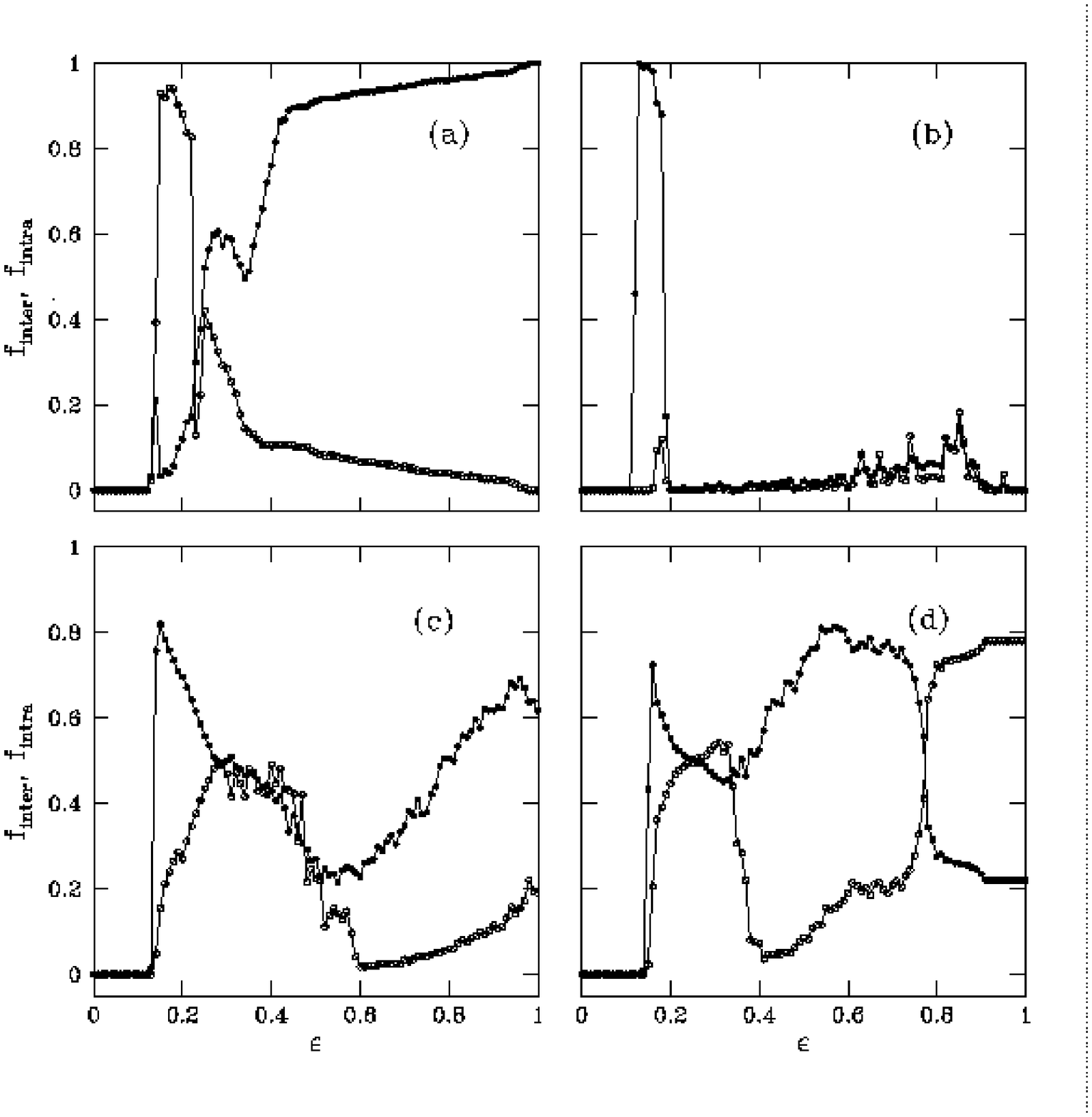}}
\caption{The fraction of intra-cluster and inter-cluster couplings,
$f_{inter}$ (solid circles) and $f_{intra}$ (open circles) are shown
as a function of the coupling
strength $\epsilon$. Figures (a) and (b) are for the one-d
coupled maps with nearest neighbor coupling ($m=1$) and
for $g(x)=x$ and $g(x)=f(x)$ respectively. Figures (c) and (d) are
for $g(x)=f(x)$ and respectively $m=5$ and $m=10$.
The figures are for $N=50$ and are obtained by averaging over 50
random initial conditions.}
\label{inter-intra-LCM-m}
\end{figure*}

For $m>1$ and $g(x) = f(x)$ we get similar kind of behaviour as for
$m=1$ with dominant driven clusters for most of the coupling strength
region, but we
do not get any ideal driven clusters. Fig.~\ref{clus-scale-m}(c) is
plotted for coupling strength $\epsilon = 0.9$ and $m=3$. 
As $m$ increases the region I showing turbulent behaviour remains unaffected,
but the mixed region II grows in
size while the region III shrinks. As $m$ increases more and more 
nodes participate in cluster formation. The driven behaviour decreases 
in strength with increasing $m$ and self-organized behaviour increases 
in strength.
For $m=10$, all nodes form one cluster for larger $\epsilon$ values 
which is obviously of the self-organized type 
(Fig.~\ref{clus-scale-m}(d)).

We have also studied the effect of size of the network on the
synchronized cluster formation. 
The phenomena of self-organized and driven behavior persists for the
largest size network that we have studied ($N=1000$). The region II 
showing self-organized or driven behavior is mostly unaffected while the
ordered regions showing driven behavior for large coupling strengths
show a small shrinking in size.

\subsection{Coupled maps on one dimensional network}
For one dimensional CMN, each node is 
connected with $m$ nearest neighbors (degree per node is
$2m$). First we consider $m = 1$, i.e each map is connected with just next
neighbors on both sides. 
Fig~.\ref{inter-intra-LCM-m}(a) and (b) show $f_{\rm intra}$ and $f_{\rm
inter}$ verses $\epsilon$ for $g(x)=x$ and $g(x)=f(x)$ respectively
and $\mu=4$, $N=50$. For $g(x)=x$,
after an initial turbulent region ($\epsilon > \epsilon_c$),
nodes form 
self-organized clusters (region II-S in Fig.~\ref{phase-scale-x}) 
and as the coupling strength increases we observe
a crossover to driven clusters.
The behaviour of clusters as well as Lyapunov
exponent graphs are similar to the scale-free network with
the coupling form $f(x)=x$.
Note that the nearest neighbor CMN with $m = 1$ is a tree 
and can be geometrically 
organized into both self-organized and driven type of clusters.

\begin{figure}
\centerline{
\includegraphics[width=9cm]{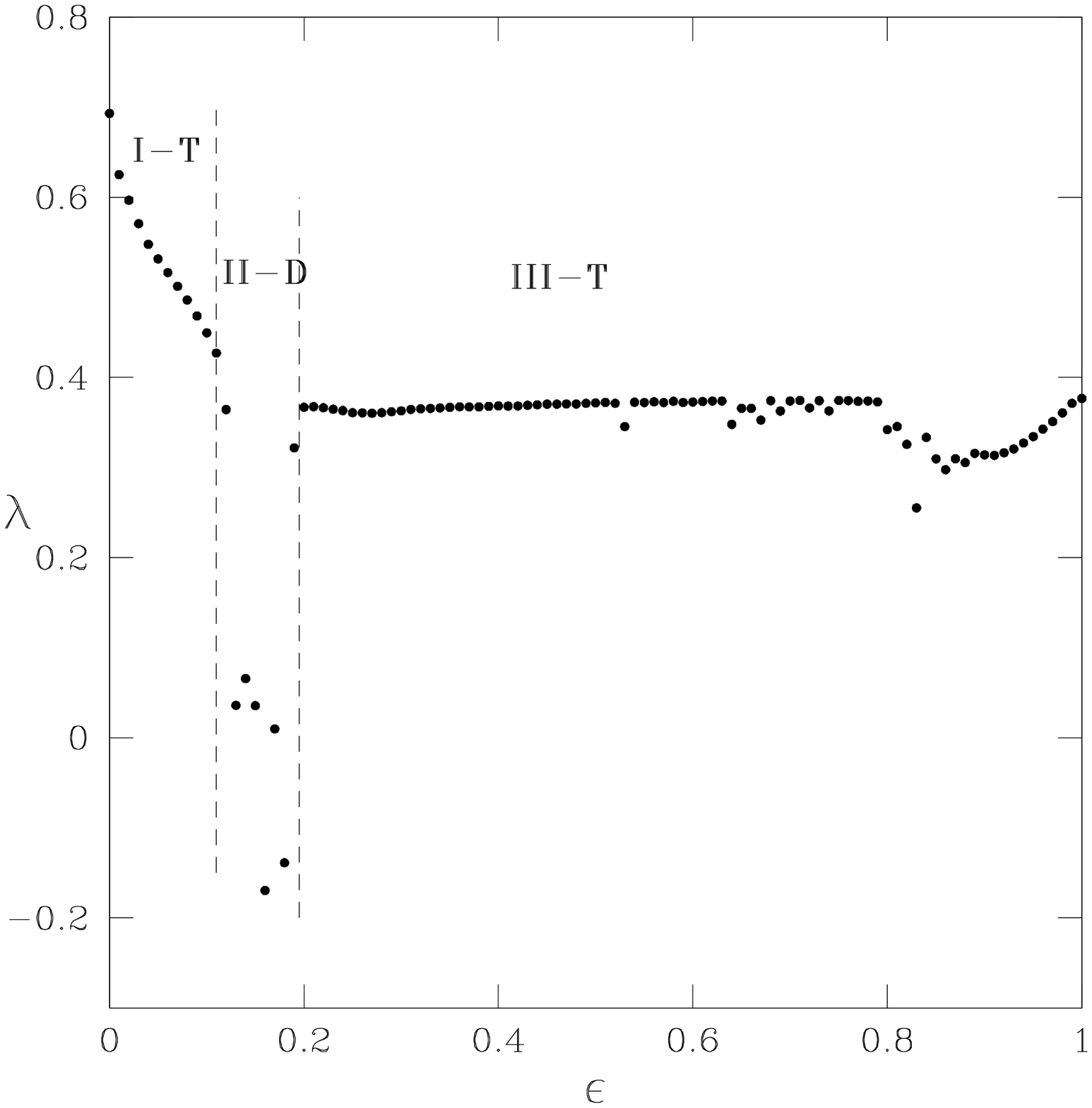}}
\caption{Largest Lyapunov exponent, $\lambda$, is plotted as a function of
$\epsilon$ for 1-d nearest neighbor coupled network for $f(x) = 4 x (1-x)$ and
$g(x)=f(x)$. Different regions are labeled as for scale-free network
(see Fig.~\ref{phase-scale-fx} and Fig.~\ref{lya-scale-fx}.)}
\label{lya-LCM-fx}
\end{figure}

For $g(x) = f(x)$ coupling and $m = 1$, there is cluster formation for 
only small coupling strength region (corresponding to region II-D of 
Fig.~\ref{phase-scale-fx})
as seen from Fig~\ref{inter-intra-LCM-m}(b).
Fig.~\ref{lya-LCM-fx} shows largest Lyapunov exponent as a function of 
$\epsilon$ for $g(x)=f(x)$ and $\mu=4.0$.
In region II-D the largest Lyapunov
exponent is positive or negative, depending
on the initial conditions and $\epsilon$ values and for the rest of the
coupling strength region 
Lyapunov exponent is positive. In region II-D synchronized
clusters of driven type are seen. For larger coupling strengths
where
the system remains in chaotic zone, clusters are rarely formed and
when formed are small in size.

\begin{figure*}
\centerline{
\includegraphics[width=18cm]{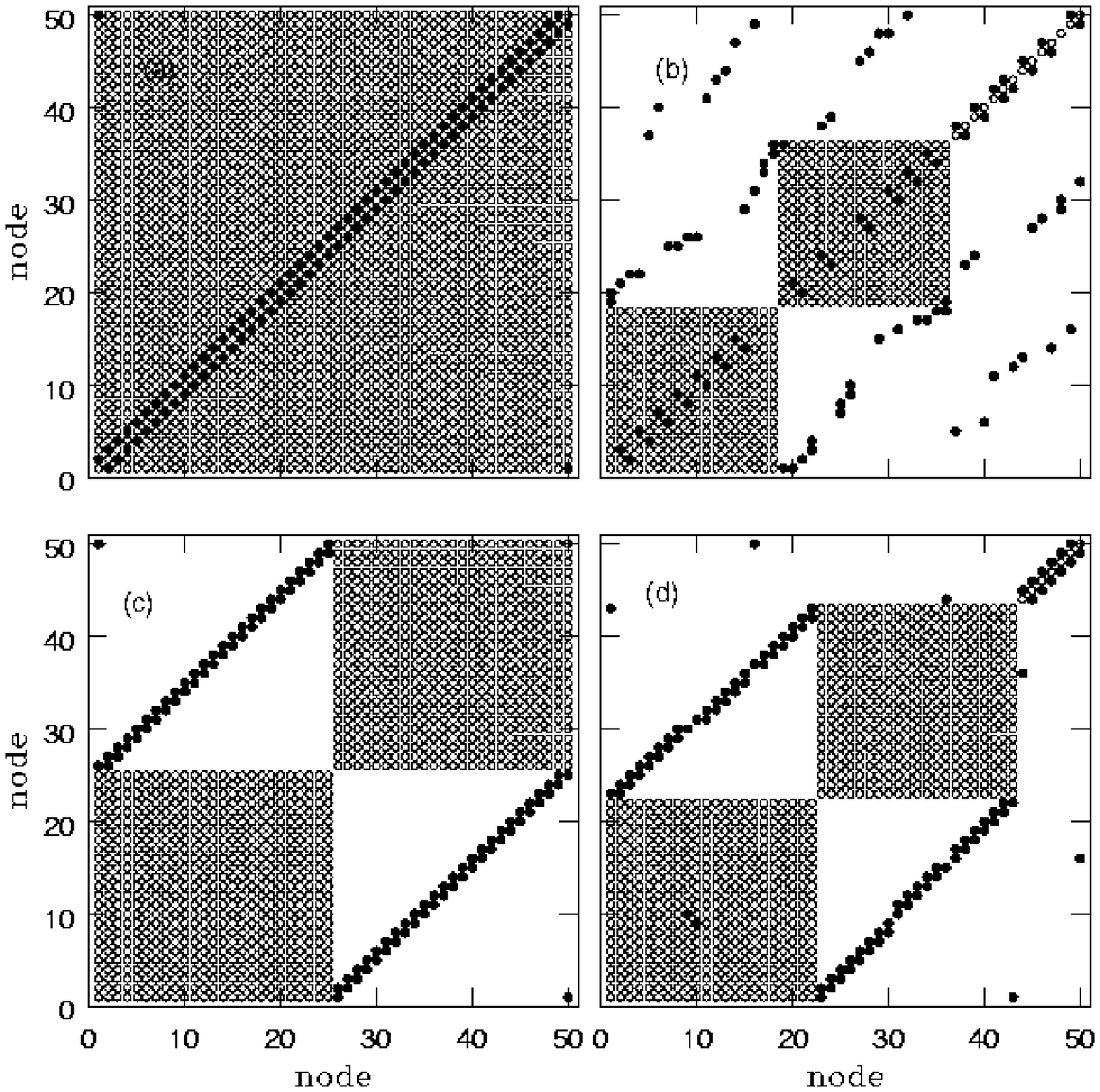}}
\caption{The figure illustrates the cluster formation for
one-d nearest neighbor network using node-node plot as in
Fig.~\ref{clus-scale-x}. (a) and (b) are
for $g(x)=x$ and $\epsilon = 0.16$,
and $\epsilon = 0.30$ respectively. (c) and
(d) are for $g(x) = f(x)$ and $\epsilon = 0.13$ and $\epsilon = 0.15$
respectively. }
\label{clus-LCM}
\end{figure*}

Fig.~\ref{clus-LCM} shows node-node plot showing synchronized
clusters. Fig.~\ref{clus-LCM}(a) shows one self-organized cluster in region 
II-S (see Fig.~\ref{phase-scale-x}) for
$g(x)=x$. In this region we also get two self-organized clusters
depending on the initial values and $\epsilon$.
Fig.~\ref{clus-LCM}(b) shows two clusters of mixed type as well as
several isolated nodes for $\epsilon$ in region III-M for $g(x)=x$. Figures
\ref{clus-LCM}(c) and~\ref{clus-LCM}(d) show driven clusters for $g(x)=f(x)$
for $\epsilon$ values in regions II-D of Fig.~\ref{lya-LCM-fx}.

We now consider the case $m>1$.
For $g(x)=x$ we observe
self-organized clusters with some inter-cluster
connections for the coupling strength region II-S
and as the coupling strength increases there is a crossover to driven
clusters. As the coupling strength increases further for $\epsilon
>0.7$ instead of forming driven clusters
(as is observed for $m=1$) nodes form one synchronized cluster.
As $m$ increases $f_{intra}$ increases and for $\epsilon >
\epsilon_c$ we observe dominance of self-organized behaviour.
For $m=5$, and for coupling strength $\epsilon > \epsilon_c (\approx
0.13)$, all nodes 
form one or two clusters. For one cluster $f_{intra}=1$ and for
two clusters intra-cluster and inter-cluster couplings
are almost equally distributed.
For very large value of coupling strength ($\epsilon > 0.7$) 
we get clusters of dominant self-organized type.
As the number of connections increases and typically becomes of the order
$N^2$ that is a globally coupled state, we get one cluster of
self-organized type.

\begin{figure*}
\centerline{\begin{tabular}{ccc}
\includegraphics[width=9cm]{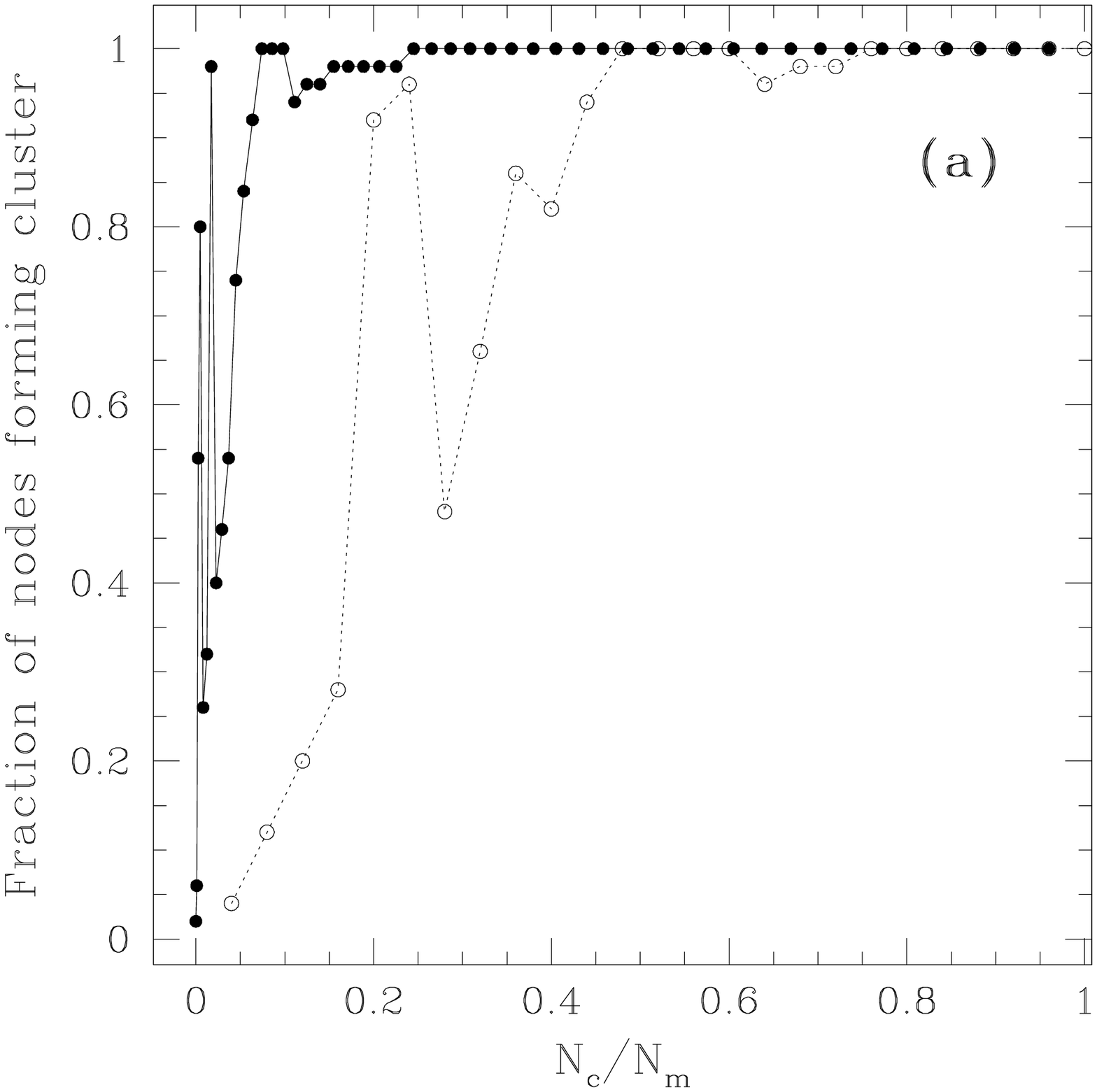} & \includegraphics[width=9cm]{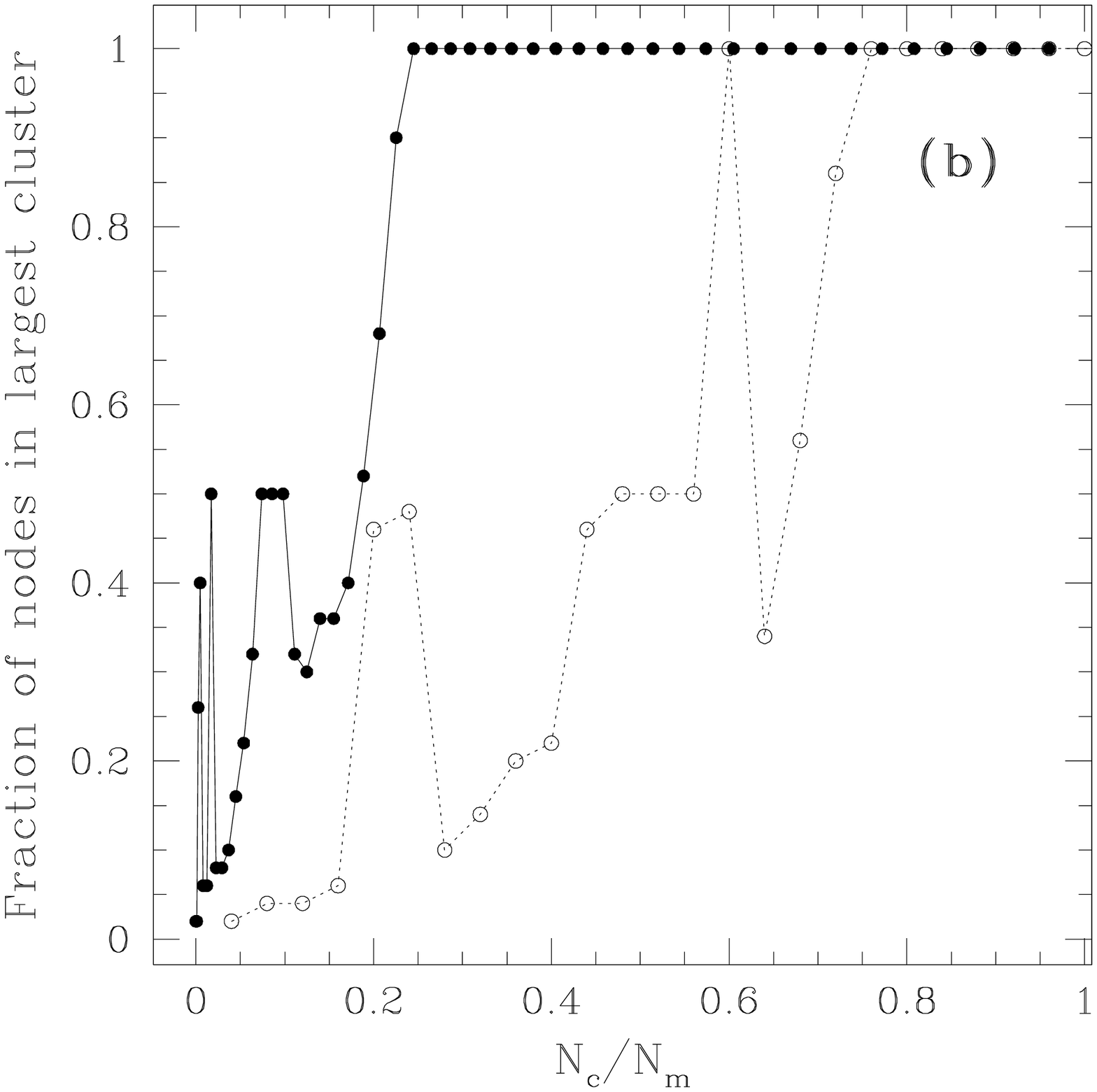}\\
\end{tabular} }
\caption{Figure (a) shows the fraction of nodes forming clusters
as a function
of the fraction of couplings $N_c/N_m$ where $N_m=N(N-1)/2$.
The figures are plotted for 1-d coupled maps with $g(x)=f(x)$
and for $\epsilon=0.49$ (closed circles) and $\epsilon=0.7$ (open
circles). The results are for $N=50$ and are obtained
by averaging over 100 random initial conditions. Figure (b)
shows the fraction of nodes in the largest
cluster as a function of
$N_c/N_m$ for $\epsilon=0.49$ (closed circles) and $\epsilon=0.7$ (open
circles). Other parameters are
same as above.}
\label{no-nodes-size-clus-LCM-fx}
\end{figure*}

For $g(x)=f(x)$ and $m>1$, we find that as the number of connections
increases for small coupling 
strength (region II-D) we get two dominant
driven phase synchronized clusters.
For large coupling strength the number of nodes
forming clusters and the sizes of clusters both increase with the increase in
number of connections in the network.
This behaviour is seen in Figs.~\ref{inter-intra-LCM-m}(c) and~(d) 
which show $f_{\rm intra}$ and $f_{\rm
inter}$ verses $\epsilon$ for $g(x)=f(x)$, $\mu=4$ and respectively for 
$m=5$ and $m=10$. Fig.~\ref{no-nodes-size-clus-LCM-fx}(a) shows the 
fraction of nodes forming
clusters as a function of the number of connections 
$N_c$ normalized with respect
to the maximum number of connections $N_m = N(N-1)/2$ for two values of
$\epsilon$. The overall increase in the number of nodes forming clusters is
clearly seen. Fig.~\ref{no-nodes-size-clus-LCM-fx}(b) shows the fraction 
of nodes in the largest
cluster as a function $N_c$ for two values of
$\epsilon$. The overall growth in the size of the clusters with $N_c$ is 
evident.

\begin{figure*}
\centerline{
\includegraphics[width=12cm]{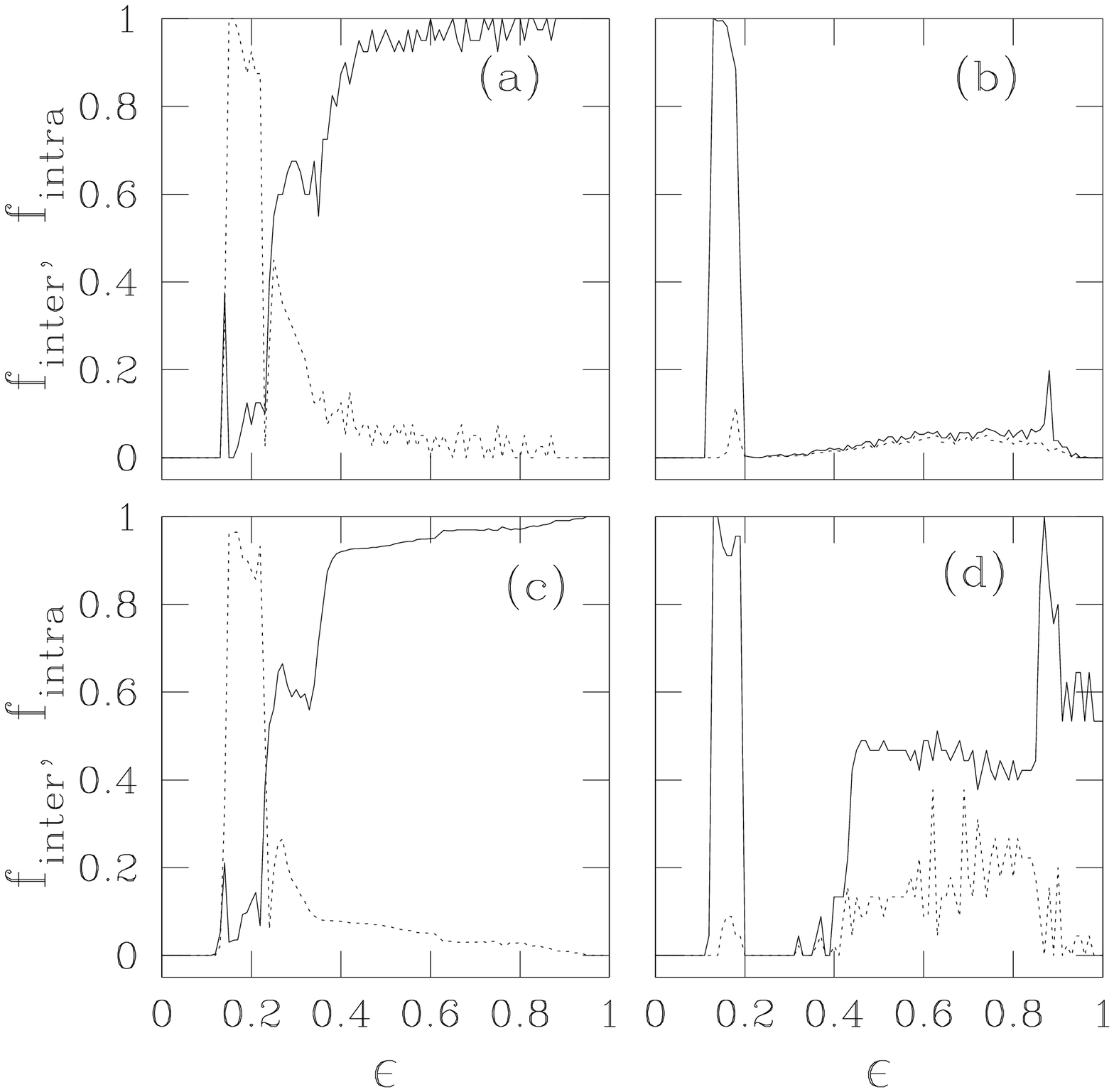}}
\caption{Frction of intra-cluster and inter-cluster couplings,
$f_{inter}$ (solid line) and $f_{intra}$ (dashed line) are shown
as a function of the coupling
strength $\epsilon$. Figures (a) and (b) are for the small world
network for $g(x)=x$ and $g(x)=f(x)$ respectively and $N=50$.
Figures (c) and (d) are
for the Caley tree with $g(x)=x$ and $g(x)=f(x)$ respectively and $N=47$.
The figures are obtained by averaging over 50
random initial conditions.
Small world networks are generated with $m=1$ and $p=0.06$ [3].
Caley trees are generated with coordination number
three [27]. }
\label{inter-intra-sated with coordination number
three [27]. }
\label{inter-intra-small-Caley}
\end{figure*}

Cluster formation with large number of connections (of
the order of $N^2$) and its dependence on coupling strength is discussed 
in Refs. \cite{CML-critical1,CML-critical2}. It is reported that for
these networks it is the coupling strength which affects
the synchronized clusters and not the number of connections. We find
that when the
number of connections is of the order of $N$ there are significant
deviations from this reported behaviour. We find that the size of the
clusters and number of nodes forming clusters increases as the number
of connections increase as discussed above. 
This behaviour approaches the
reported behaviour as the number of connections increases and becomes
of the order of $N^2$.

\subsection{Coupled maps on Small world network} 
Small world networks are constructed using the following algorithm 
by Watts and Strogatz \cite{Watts}. Starting with a one-dimension ring
lattice of $N$ nodes in
which every node is connected to its nearest $k$ neighbors ($k/2=m$
on either side), we randomly rewire each connection of the 
lattice with probability $p$ such that self-loops and multiple
connections are excluded. Thus, $p=0$ gives a regular network and
$p=1$ gives a random network. Here we present results for
$N = 50$ and $m=1$. 
Figs.~\ref{inter-intra-small-Caley}(a)
and~\ref{inter-intra-small-Caley}(b) plot $f_{\rm intra}$ and $f_{\rm
inter}$ for $g(x)=x$ and $g(x)=f(x)$ respectively as a function of
$\epsilon$ for $\mu=4$. We find that
for $g(x) = x$, the behaviour is very similar to that for the scale free
networks and one-d lattice. We get self-organized clusters for $\epsilon >
\epsilon_c$ and there is a crossover to driven behavior as epsilon
increases (Fig.~\ref{inter-intra-small-Caley}(a) ). 
But for 
$g(x) = f(x)$, nodes form clusters only for region II-D of 
coupling strength 
and there is almost no cluster formation for larger values of
$\epsilon$ ( Fig.~\ref{inter-intra-small-Caley}(b) ). This behaviour
changes as $k$ increases and we
observe some clusters for large $\epsilon$ values
also. 
Figure~\ref{clus-diff}(a) shows node-node plot of clusters for 
$\epsilon=0.45$, $m=1$ and $g(x)=x$ showing dominant driven clusters.

\subsection{Coupled maps on Caley Tree}
We generate a Caley tree using the algorithm given in Ref. \cite{Tree}.
Starting with three branches at the first level, we split each branch 
into two at subsequent levels.
For $g(x)=x$, the behaviour is similar to all other networks with the same
number of connections (Fig.~\ref{inter-intra-small-Caley}(c)). 
Figure~\ref{clus-diff}(b) shows
node-node plot of two ideal driven phase synchronized clusters for
$\epsilon=0.92$, $\bar{k} = 2$, $N=47$ and $g(x)=x$.
For $g(x)=f(x)$ all nodes form driven clusters for 
region II-D, and for larger coupling strengths
about 40\% of nodes form clusters of driven types
(Fig.~\ref{inter-intra-small-Caley})(d)).

\subsection{Coupled maps on higher dimensional lattices}
\begin{figure*}
\centerline{
\includegraphics[width=17cm]{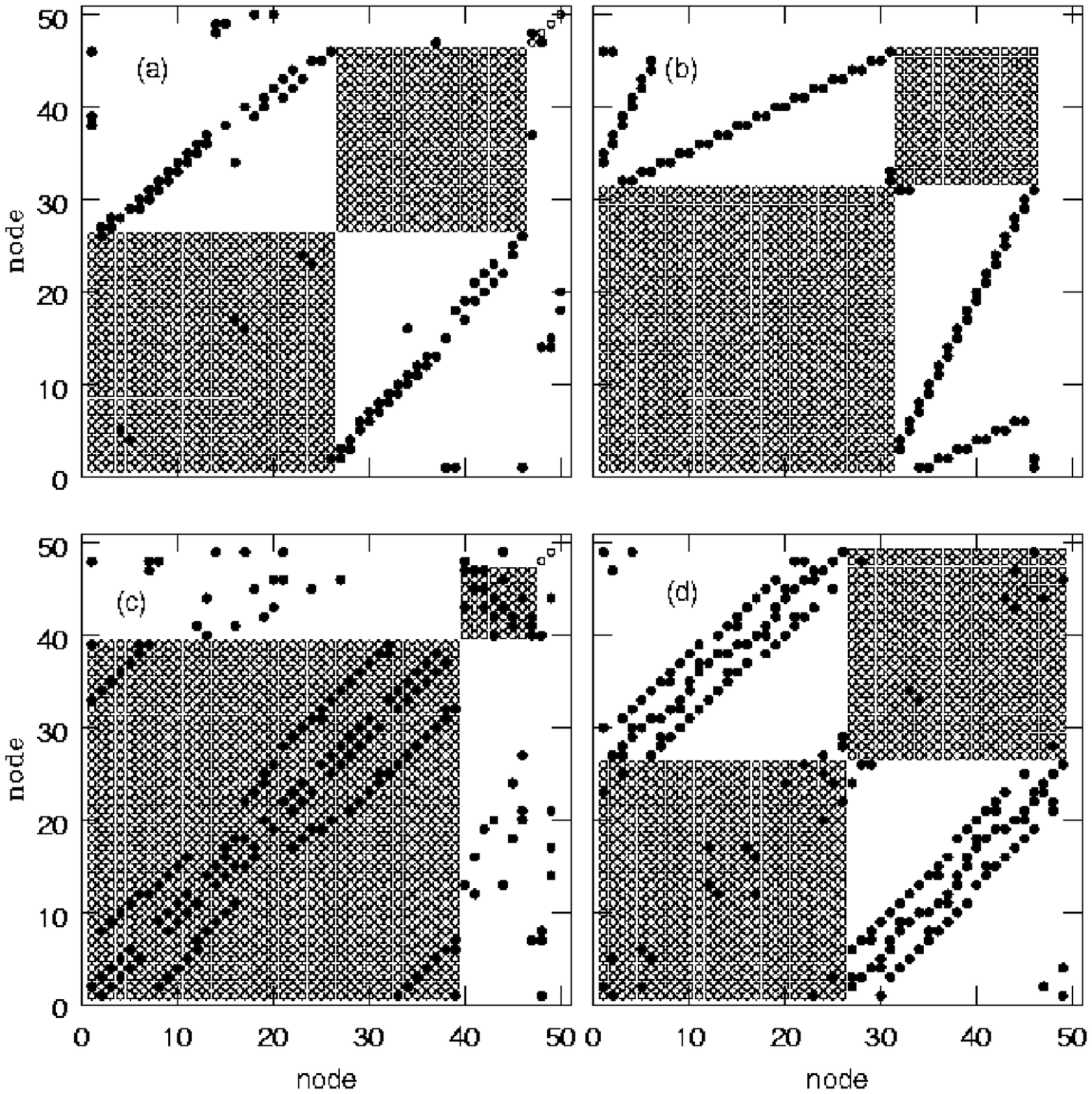} }
\caption{Figure illustrates the cluster formation for
different networks as node vs node plot as in
Fig~\ref{clus-scale-x}. (a) is plotted for small world network with
$\epsilon=0.45$, $N=50$ and $g(x)=x$. (b) is plotted for Caley tree with
$\epsilon=0.92$, $N=47$ and $g(x)=x$.
(c) and (d) are plotted for 2-d lattice ($N=49$) with $\epsilon=0.21$, $g(x)=x$
and $\epsilon=0.19, g(x)=f(x)$ respectively. }
\label{clus-diff}
\end{figure*}

\begin{figure}
\centerline{
\includegraphics[width=4.5cm]{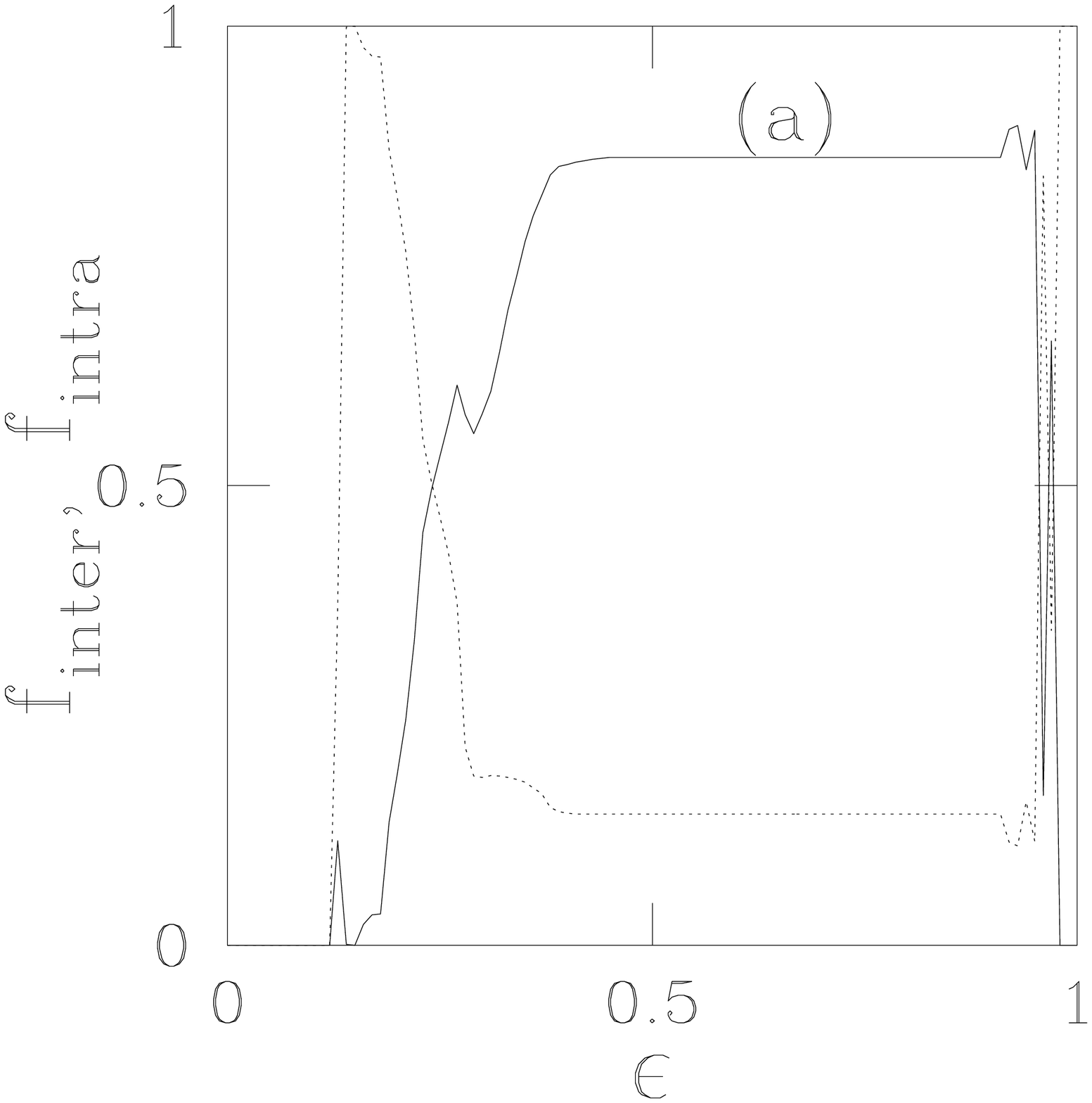}
\includegraphics[width=4.5cm]{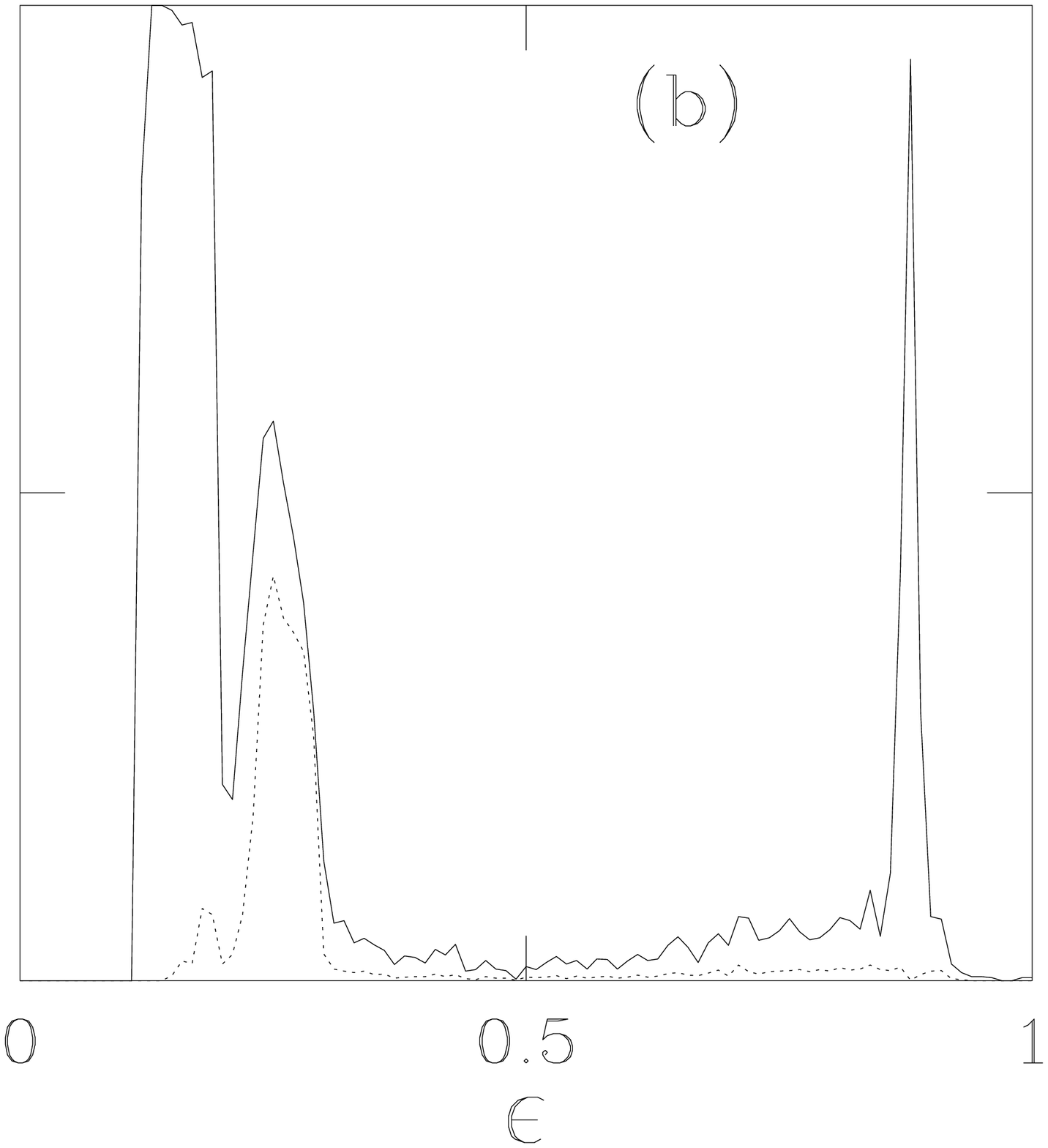}}
\caption{The fraction of intra-cluster and inter-cluster couplings,
$f_{inter}$ (solid line) and $f_{intra}$ (dashed line) are shown
as a function of the coupling
strength $\epsilon$ for two-d lattice. Figures (a) and (b) are
for $g(x)=x$ and $g(x)=f(x)$ respectively. The figures are for $N=49$ and
are obtained by averaging over 50 random initial conditions. }
\label{inter-intra-2d}
\end{figure}

Coupled maps on higher dimensional lattices also form synchronized
clusters. First
we give the result for two-d square lattices. 
Figs.~\ref{inter-intra-2d}(a)
and~\ref{inter-intra-2d}(b) plot $f_{\rm intra}$ and $f_{\rm
inter}$ for $g(x)=x$ and $g(x)=f(x)$ respectively as a function of
$\epsilon$ for $\mu=4$.
For $g(x)=x$
the cluster formation is similar to other networks described earlier
except for very large $\epsilon$ close to one where we get a single
self-organized cluster.  
For $g(x)=f(x)$ cluster formation is similar to that in  one-d
networks with nearest and next nearest neighbor couplings. 
In small coupling strength region II-D
(see Figure~\ref{phase-scale-fx}), nodes form two clusters of driven type
and for large coupling strength also driven clusters are observed
with 25-30\% nodes showing synchronized behaviour 
(Figure.~\ref{inter-intra-2d}(b)). Figures~\ref{clus-diff}(c)
and~(d) show node-node plot of
self-organized behaviour for $g(x)=x$ and dominant driven behaviour
for $g(x)=f(x)$. 

Coupled maps on three-d cubic lattice (degree per node is six)
for $g(x)=x$ show clusters similar to the other networks discussed
earlier. 
For $g(x)=f(x)$, nodes form driven type of clusters
at small coupling strength (region II-D) and mainly 
we observe three clusters. For large
coupling strengths also nodes form driven clusters and the nodes
participating in cluster formation is now much larger than the two-d case. 

\subsection{Coupled maps on random network}
Random networks are constructed by connecting each pair of nodes with 
probability $p$.
First consider the case where the average degree per node is two. 
For linear coupling $g(x)=x$ cluster formation is the same as for other
networks with same average degree. For $g(x)=f(x)$ 
driven type clusters are
observed in region II-D and no significant cluster
formation is observed for larger coupling strengths. This behaviour is 
similar to one-d network with $k=2$ but different from the
corresponding scale free network.
For coupled maps on random networks with average degree per node equal
to four and $g(x)=f(x)$, clusters with dominant driven behaviour are
observed for all $\epsilon > \epsilon_c$.

\subsection{Examples of self-organized and driven behavior}
There are several examples of self-organized and driven behaviour in
naturally occurring systems. 
An important
example in physics that includes both self-organized and driven
behavior, is the nearest neighbor Ising model treated 
using Kawasaki dynamics.
As the strength of the Ising interaction between spins changes sign
from positive to negative
there is a change of phase from a ferromagnetic (self-organized) to an
antiferromagnetic (driven) behavior. In the antiferromagnetic state,
i.e. driven behavior,
the lattice spits into two sub-lattices with
only inter-cluster interactions and no intra-cluster interactions.

Several other examples are discussed in Ref.~\cite{sarika-REA1}.

\section{Universal features}
We have discussed the behavior of evolution of coupled dynamical
elements on different networks. For small coupling strengths unto a
critical value $\epsilon_c$
nodes show turbulent behaviour and for
$\epsilon > \epsilon_c$ they form interesting phase-synchronized
clusters. The
critical value $\epsilon_c$  depends on the type of network, the type
of coupling function and parameter $\mu$.
We find that clusters are formed
because of intra-cluster connections and/or inter-cluster 
connections depending on the network, 
the type of coupling and the coupling strength. 

The linear coupling of type $g(x) = 
x$ shows both types of clusters (self-organized and driven) and show a 
behaviour which is universal for all the networks that we have studied.
For networks, with number of connections of the order of $N$,
initially for a small range of coupling strength (region II-S in
Figure~1) nodes form 
self-organized clusters and
as coupling strength increases there is a crossover and reorganization
of nodes to driven clusters. This behaviour is observed for all the
networks that we have studied i.e. scale-free network,
random network, small world 
network, Caley tree network, 1-d nearest neighbor coupled network, 
1-d next to next neighbor coupled network, 2-d network.
As the number of connections, $N_c$, increases, the driven behaviour for large
coupling strengths is suppressed and self-organized behaviour starts 
dominating. As $N_c$ increases further a clear 
identification of the two mechanisms,
self-organized and driven, becomes more and more difficult.
As $N_c$ becomes of the order of $N^2$,
for very large coupling strengths we observe one
spanning cluster of self-organized type.

For coupling function $g(x)=f(x)$ the formation of clusters
and nature of clusters both depend on the type of network.
Initially for a small range of coupling strength values (region II-D in
Fig.~\ref{phase-scale-fx}) nodes form driven clusters for all
networks but as coupling strength
increases cluster formation and size of clusters both depend on
the type of network.
For networks with average connectivity per node equal to two, we find that
for large coupling strengths number of nodes forming clusters and
size of clusters both are small. 
For large coupling strengths where
largest Lyapunov exponent is positive,
less than 50\% nodes form clusters for scale-free network and
Caley tree. For other types of networks cluster formation is not significant.
As the total number of connections increases from $N_c \sim N$ to 
$N_c \sim N^2$, the number of nodes forming clusters as well as the size
of clusters increase. Again as in the case of $g(x)=x$ for very large
coupling strengths one large spanning cluster is observed.

It is interesting to note that nodes can form two or more
stationary clusters
even though coupled dynamics is in the chaotic regime.
Here, a stationary cluster means that constituents of the cluster are unique
and once nodes form a stationary cluster they belong to that cluster
forever, that is the structure of the cluster 
does not depend on time.
For $\mu=4$, we get
stationary clusters when the nodes form two clusters, but they are not 
stationary when they form three or more clusters. Note
that three or more stationary clusters can be formed for $\mu<4$.

For $\mu=4$ if the largest Lyapunov exponent is negative, the variables show
periodic behaviour with even period. For $\mu<4$ the periodic
behaviour can have both odd and even periods.

We observe ideal behaviour of both the types that is all the
nodes forming driven clusters ($f_{inter}=1$) or all the nodes forming
self-organized clusters ($f_{inter}=1$). In most cases where
ideal behaviour is observed, the largest Lyapunov
exponent is negative or zero giving stable clusters. However, in
some cases ideal behaviour is also observed in the chaotic region.

\section{Circle map}
We have studied cluster formation by considering circle map as
defining the local dynamics, given by
\[ f(x) = x + \omega + (k/2\pi) sin(2\pi x) ,\,  \pmod{1} \]   
Due to the modulo condition, instead of using the variable $x_t$, we use a
function of $x_t$ such as $\sin(\pi x_t)$ satisfying periodic boundary
conditions to decide the location of maxima and
minima which are used to determine the phase synchronization of two nodes
(Eq.~(\ref{phase-dist})).
With circle map also we observe formation of
clusters with the time evolution starting from initial random conditions.
 Here we discuss the results with the
parameters of the circle map in the chaotic region ($\omega=0.44$ and
$k=6$). For linear
coupling $g(x)=x$ and scale-free networks with $m=1$,
for small coupling strength nodes evolve chaotically with
no cluster formation. As coupling strength increases nodes form
clusters for $0.21 < \epsilon < 0.25$. In most of this region the
nodes form two cluster and these clusters are mainly of
the driven type except in the initial part, $\epsilon \approx 0.21$,
where self-organized clusters can be observed. 
As the coupling strength increases nodes 
behave in a turbulent manner and after $\epsilon > 0.60$ nodes form
clusters of dominant driven type. Here the number of nodes forming
clusters and the sizes of clusters, both are small.
For the one dimensional linearly coupled networks, for linear coupling
the nodes form phase synchronized clusters for coupling strength
region $0.21 < \epsilon < 0.25$. The clusters are mainly of the driven
type except in the initial part, $ \epsilon \approx
0.21$, where they are of the self-organized type. For
large coupling strength they do not show any cluster formation.

For $g(x)=f(x)$ we found very negligible cluster formation for the
entire range of the coupling strength for both scale free and one-d
network. However, as $m$ increases the nodes
form phase synchronized clusters for $\epsilon$ larger than some 
critical $\epsilon_c$.

For the circle map the normalization factor $(1-\epsilon)$ in the first term of
Eq.~(\ref{coupleddyn}) is not necessary and the following modified model can 
also be considered.
\begin{equation}
x^{i}_{t + 1} = f( x^{i}_t ) + \frac{\epsilon}{k_i}
\sum_{j=1}^N C_{ij} g( x^{j}_t ), \; \; \pmod{1}.
\label{coupleddyn2}
\end{equation}
We now discuss the synchronized cluster formation for the same
parameter values as above  ($\omega=0.44$ and
$k=6$) for this modified model.
For linear coupling, clusters are formed only for $0.02 < \epsilon <0.17$
with dominant self-organizied behaviour for most of the range
except near $\epsilon \approx 0.17$ where the behaviour is of doninant
driven type. For the scale free networks ($m=1$) we have ordered states while
for the one-d networks we have partially ordered states.  For nonlinear
coupling, the clusters are formed for $0.0 < \epsilon <0.09$. The
scale-free networks show mostly mixed type clusters while one-d
networks show dominant self-organized clusters. There is no
cluster formation for larger coupling strengths for both linear and
nonlinear coupling. However, as for the logistic map, synchronized
clusters are obseved for large $\epsilon$ as the number of connections 
inceases.

\section{Conclusion and Discussion}
We have studied the properties of coupled dynamical elements on
different types of networks. We find that
in the course of time evolution they show phase synchronized 
cluster formation. 
We have mainly studied networks with small number of connections
$(N_c \sim N)$ because a large number of
natural systems fall under this category
of small connections. More importantly,
with small number of connections,
it is easy to identify the relation between the dynamical
evolution, the cluster formation and the geometry of networks.
We have studied the mechanism of cluster formation
as well as behaviour of individual nodes either forming clusters
or evolving independently. We use the logistic map for the local
dynamics and the two types of couplings.
Depending upon the type of coupling, the regions for cluster 
formation and the behaviour of phase synchronization vary.
We have identified two mechanism of cluster formation, 
self-organized and driven phase synchronization.

By considering the number of inter- and intra-cluster couplings we can 
identify phase synchronized clusters with dominant self-organized 
behavior (S), dominant
driven behavior (D) and mixed behavior (M) where both mechanisms
contribute. We have also observed ideal clusters of both
self-organized and driven type. In most cases where
ideal behaviour is observed, the largest Lyapunov
exponent is negative or zero giving stable clusters with periodic
evolution. However, in
some cases ideal behaviour is also observed in the chaotic region.
In most of the cases when synchronized clusters are formed there are some
isolated nodes which do not belong to any cluster. More interestingly
there are some {\it floating} nodes which show an intermittent behavior
between an independent evolution and an evolution synchronized with some
cluster. The time spent by a floating node in the synchronized cluster 
shows an exponential distribution.

By defining different states of the dynamical system using the number and type
of clusters, we consider the phase-diagram in the $\mu - \epsilon$
plane for the local dynamics governed by the logistic map.
When the local dynamics is in the chaotic region,
for small coupling strengths we observe turbulent behaviour. There is 
a critical value $\epsilon_c$ above which phase synchronized clusters
are observed.
For $g(x) = x$ type of coupling, self-organized clusters are formed when
the strength of the coupling is small.
As the coupling strength increases there is a crossover from
the self-organized to the driven behavior which also involves
reorganization of nodes into different clusters. 
This behaviour is
almost independent of the type of networks.
For non linear coupling of type $g(x) = f(x)$,  for small coupling 
strength
phase synchronized clusters of driven type are formed, but for large coupling
strength number of nodes forming cluster as well as size of cluster
both are very small and almost negligible for many network. Only for
scale-free networks and Caley tree show some cluster formation for
large coupling strengths.

As the number of connections increases, most of the clusters become of 
the mixed type where both the mechanisms contribute. We find that in
general, the self-organized behaviour
is strengthened and also the number of nodes forming clusters as 
well as the size of clusters increase. As the number of connections
become of the order of $N^2$, self-organized behaviour with a single
spanning cluster is observed for $\epsilon$ larger than some value.

It is interesting to consider the dynamical origin of the
self-organized and driven
phase synchronization. A clue can be obtained by considering small
networks of two or three nodes and also tailormade large networks such 
as globally coupled networks and complete bipartite networks. These
are considered in Ref.~\cite{pre2}. These studies reveal that  the
intra-cluster coupling term between the two varibles $x^i$ and $x^j$, 
adds a decay term to the
dynamics of the difference variable $x^d = x^i - x^j$ leading to 
synchronization of these two variables. On the other hand, the
inter-cluster coupling term between the two varibles $x^i$ and $x^j$, 
cancels out in the dynamics of $x^d$ and the two variables now
belong to different synchronized clusters.

In this paper we have discussed the numerical results of phase
synchronization on CMN and the two mechanisms of cluster formation. In another paper we discuss the stabilty analysis of
synchronized clusters in some simple networks that illustrate the two 
mechanisms of synchronization \cite{pre2}.

\appendix
\section{}
Here we show that the definition (\ref{phase-dist}) of phase
distance $d_{ij}$
between two nodes $i$ and $j$ satisfies metric properties. Let ${\cal
N}_i$ denote the set of minima of the variable $x^i_t$ in a time
interval $T$. The phase distance satisfies the following metric
properties. \\
(A) $d_{ij} = d_{ji}$. \\
(B) $d_{ij} =0$ only if ${\cal N}_i = {\cal N}_j$. \\
(C) Triangle inequality:  Consider
three nodes $i$, $j$ and $k$. Denoting the number of elements of a set 
by $|.|$, let, \\
(1) $a = |{\cal N}_i \cap {\cal N}_j \cap {\cal N}_k$. \\
(2) $b = |{\cal N}_i \cap {\cal N}_k| -a$.\\
(3) $c = |{\cal N}_j \cap {\cal N}_k| -a$.\\
(4) $d = |{\cal N}_i \cap {\cal N}_j| -a$.\\
(5) $e = |{\cal N}_i| -b -d -a$. \\
(6) $f = |{\cal N}_j| -c -d -a$. \\
(7) $g = |{\cal N}_k| -b -c -a$. \\
We have
\begin{eqnarray}
n_{ik} & = & a + b \nonumber \\
n_{jk} & = & a + c \nonumber \\
n_{ij} & = & a + d \nonumber \\
n_i & = & a + b + d + e \nonumber \\
n_j & = & a + c + d + f \nonumber \\
n_k & = & a + b + c + g \nonumber 
\end{eqnarray}
Consider the combination
\begin{equation}
d_{ik} + d_{jk} - d_{ij} = 1 - X
\end{equation}
where
\[ X = \frac{n_{ik}}{{\rm max}(n_i,n_k)} + \frac{n_{jk}}{{\rm
max}(n_j,n_k)} - \frac{n_{ij}}{{\rm max}(n_i,n_j)} \]
The triangle inequality is proved if $X \leq 1$. Consider the following 
three general cases.
\begin{itemize}
\item[Case A.] $n_i \leq n_j \leq n_k$: \\
\begin{eqnarray}
X & = & \frac{a+b}{n_k} + \frac{a+c}{n_k} - \frac{a+d}{n_j} 
\nonumber \\
& \leq & \frac{a+b+c-d}{n_k}
\nonumber \\
& \leq & 1
\end{eqnarray}

\item[Case B.] $n_i \leq n_k \leq n_j$: \\
\begin{eqnarray}
X & = & \frac{a+b}{n_k} + \frac{a+c}{n_j} - \frac{a+d}{n_j} 
\nonumber \\
& \leq & \frac{a+b+c}{n_k}
\nonumber \\
& \leq & 1
\end{eqnarray}

\item[Case C.] $n_k \leq n_i \leq n_j$: \\
\begin{eqnarray}
X & = & \frac{a+b}{n_i} + \frac{a+c}{n_j} - \frac{a+d}{n_j} 
\nonumber \\
& \leq & \frac{a+b+c}{n_i}
\nonumber \\
& \leq & \frac{a+b+c}{n_k}
\nonumber \\
& \leq & 1
\end{eqnarray}

\end{itemize}
This proves the triangle inequality.


\begin{thebibliography}{99}	

\bibitem{Strogatz} S. H. Strogatz, Nature, {\bf 410}, 268 (2001) and
references theirin.
\bibitem{rev-Barabasi} R. Albert and A. L. Barab\"asi,
Rev. Mod. Phys. {\bf 74}, 47 (2002) and references theirin.
\bibitem{Watts} D. J. Watts and S. H. Strogatz, Nature (London) {\bf 393},
440 (1998).
\bibitem{scalefree} A. -L. Barab\"asi, R. Albert, Science, {\bf 286}, 509 (1999).

\bibitem{koch} C. Koch and G. Laurent, Science {\bf 284}, 96 (1999).
\bibitem{social} S. Wasserman and K. Faust, Social Network Analysis, 
Cambridge Univ. Press, Cambridge 1994.
\bibitem{www} R. Albert, H. Jeong, A. -L. Barab\"asi, Nature {\bf 401},
130 (1999); R. Albert, H. Jeong, A. -L. Barab\"asi, Nature {\bf 406},      
378 (2000). 
\bibitem{metabolic} H. Jeong, B. Tomber, R. Albert, Z. N. Oltvai,
A. -L. Barab\"asi, Nature, {\bf 407}, 651 (2000).
\bibitem{food} Richard J. Williams, Nea D. Martinez, Nature, 
{\bf 404}, 180 (2000).
\bibitem{citation} S. Render, Euro. Phys. J. B. {\bf 4}, 131 (1998); 
M. E. J. Newman, Phys. Rev. E {\bf 64} 016131 (2001).

\bibitem{LCM1} K. Kaneko, Prog. Theo. Phys. {\bf 72}
No. 3, 480 (1984).
\bibitem{LCM2} K. Kaneko, Physica D, {\bf 34}, 1 (1989), and 
references therein.

\bibitem{GCM1} K. Kaneko, Phys. Rev. Lett. {\bf 65}, 1391 (1990);
Physica D, {\bf 41}, 137 ( 1990 ); Physica D, {\bf 124}, 322 (1998).
\bibitem{GCM2} D. H. Zanette and A. S. Mikhailov, Phys. Rev. E {\bf 57},
276 (1998).
\bibitem{GCM3} N. J. Balmforth, A. Jacobson and A. Provenzale, 
CHAOS {\bf 9}, 738 (1999).
\bibitem{GCM4} O. Popovych, Y. Maistenko and E. Mosekilde,
Phys. Lett. A, {\bf 302}, 171 (2002).
\bibitem{GCM5} G. I. Menon, S. Sinha, P. Ray, (Europhy. Let.) 
arXiv:cond-mat/0208243.
\bibitem{GCM6} O. Popovych, A. Pikovsky and Y. Maistrenko, Physica D,
{\bf 168-169}, 106 (2002).
\bibitem{GCM-phasedia} G. Abramson and D. H. Zanette, PRE, {\bf 58}, 4454
(1998).

\bibitem{HChate-phase} A. Lema\^itre and H. Chate, Phys. Rev. Lett. {\bf 82},
1140 (1999). 

\bibitem{Kaneko-2000} N. B. Ouchi and K Kaneko, Chaos {\bf 10}, 359
(2000).
\bibitem{H-Chate} H. Chate and P. Manneville, Europhys. Lett.
{\bf 17}, 291 (1992); Chaos 2 {\bf 3}, 307 (1992).
\bibitem{P-Gade} P. M. Gade, Phys. Rev. E {\bf 54}, 64 (1996).

\bibitem{random-net1} C. Susanna, Manrubia and A. S. Mikhailov,
PRE, {\bf 60}, 1579 (1999).
\bibitem{random-net2} Y Zhang, G. Hu, H. A. Cerderia, S. Chen,
T. Barun and Y. Yao, PRE {\bf 63}, 63 (2001).
\bibitem{random-net3} S. Sinha, PRE {\bf 66} 016209 (2002).

\bibitem{Tree} P. M. Gade and H. A. Cerderia, R. Ramaswamy, PRE {\bf52},
2478 (1995).
\bibitem{Small-1} 
P. M. Gade and C.-K. Hu, Phys. Rev. E  {\bf 62}, 6409 (2000);
\bibitem{Small-2} H. Hong, M. Y. Choi and B. J. Kim, Phys. Rev. E {\bf 65}
026139 (2002); Phys. Rev. E {\bf 65} 047104 (2002).
\bibitem{Small-3} T. Nishikawa, A. E. Motter, Y. C. Lai and
F. C. Hoppensteadt, Phys. Rev. Lett. {\bf 91}, 014101 (2003). 

\bibitem{CHAOS-2002} N. J. Balmforth, A. Provenzale and R. Sassi,
CHAOS {\bf 12}, 719 (2002).

\bibitem{CML-circle} S. E. de S. Pinto and R. L. Viana, Phys.
Rev. E {\bf 61} 5154 (2000).

\bibitem{CML-stability1} H. Fujisaka and T. Yamada, Prog. Theo. Phys. 
{\bf 69} 32 (1983).
\bibitem{CML-stability2} M. Ding and W. Yang, Phys. Rev. E {\bf 56},
4009 (1997).
\bibitem{CML-stability3} M. Soins and S. Zhou, Physica D {\bf 165},
12 (2002).
\bibitem{network-stability4}  J. Jost and M. P. Joy, Phys. Rev. E {\bf 65},
016201 (2002).
\bibitem{bi-partite1} Y. L. Maistrenko, V. L. Maistrenko, O. Popovych
and E. Mosekilde, Phys. Rev. E, {\bf 60} 2817 (1999).

\bibitem{REA-gade} R. E. Amritkar, P. M. Gade and A. D. Gangal, Phys.
Rev. A {\bf 44}, R3407 (1991).

\bibitem{CML-REA} R. E. Amritkar, Physica A, {\bf 224},
382 (1996).
\bibitem{CML-other1} E. Olbrich, R. Hegger and H. Kantz, Phys. Rev. Lett.
{\bf 84} 2132 (2000).
\bibitem{CML-other2} P. Garc\"{i}a, A. Parravano, M. G. Cosenza,
J. Jim\"{e}nez and A. Marcano, Phys. Rev. E {\bf 65} 045201-1 (2002).
\bibitem{CML-other3} L. Cisneros, J. Jim\"{e}nez, M. G. Cosenza
and A. Parravano, Phys. Rev. E {\bf 65} 045204-1 (2002).

\bibitem{ex-GCM}H. Aref, Ann. Rev. Fluid Mech. {\bf 15}, 345 (1983)
\bibitem{ex-optical} G. Perez, C. Pando-Lambruschini, S. Sinha and 
H. A. Cerderia, Phys. Rev. A, {\bf 45} 5469 (1992).
\bibitem{ex-Rayleigh} E. A. Jackson and A. Kodogeorgiou, Phys. Lett. A, 
{\bf 168}, 270 (1992).
\bibitem{ex-convection} T. Yanagita and K. Kaneko, Phys. Lett. A, 
{\bf 175}, 415 (1993).
\bibitem{ex-stock} C. Tsallis, A. M. C. de Souza and E. M. F. Curado,
Chaos, Solitons and Fractals, {\bf 6}, 561 (1995).
\bibitem{ex-eco} R. J. Hendry, J. M. McGlade and J. Weiner, 
Ecological Modeling, {\bf 84}, 81 (1996).
\bibitem{ex-CML} S. Sinha and W. L. Ditto, {\it Phys. Rev. Lett.}, 
{\bf 81}, 2156 (1998).
\bibitem{ex-soliton} K. Aoki and N. Mugibayashi, Phys. Lett. A, {\bf 128},
349 (1998).
\bibitem{ex-celegans} F. A. Bignone, Theor. Comp. Sci., {\bf 217} 157
(1999).

\bibitem{sarika-REA1} S. Jalan, R. E. Amritkar, Phys. Rev. Lett. 
{\bf 90}, 014101 (2003).

\bibitem{book-syn} A. Pikovsky, M. Rosenblum and J. Kurth, 
{\it Synchronization : A universal concept in nonlinear dynamics} 
(Cambridge University Press, 2001). 
\bibitem{phy-report} S. Boccaletti, J. Kurth, G. Osipov, D.L. Valladares 
and C. S. Zhou, Physics Report, {\bf 366} (1-2), (2002), The 
synchronization of chaotic systems.
\bibitem{syn-coup} Y. Zhang, G. Hu, H. A. Cerdeira, S. Chen, T. Braun
and Y. Yao, Phys. Rev. E {\bf 63} 026211-1 (2001).
 
\bibitem{phase1} M. G. Rosenblum, A. S. Pikovsky, and J.
Kurth, PRL, {\bf 76}, 1804 (1996); W. Wang, Z. Liu, Bambi Hu, Phys. rev. Lett.
{\bf 84}, 2610 (2000). 
\bibitem{phase2} S. C. Manrubia and A. S. Mikhailov, Europhys. Lett. 
{\bf 53} (4), 451 (2001).
\bibitem{phase3} N. B. Janson, A. G. Balanov, V. S. Anishchenko and
P. V. E. McClintock, Phys. Rev. Lett. {\bf 86} 1749 (2001).
\bibitem{phase4} H. L. Yang, Phys. Rev. Lett. {\bf 64} 026206-1 (2001).

\bibitem{syn} F. S. de San Roman, S. Boccaletti, D. Maza and H. Mancini,
 Phys. Rev. Lett. {\bf 81}, 3639 (1998).

\bibitem{note1} In an earlier paper \cite{sarika-REA1} we used the
definition of phase
distance as $d_{ij}=1-2\nu_{ij} / (\nu_i+\nu_j)$. However, this
earlier definition does not satisfy the triangle inequality, while the
definition given in this paper satisfies the same.

\bibitem{different-region} S. C. Manrubia and A. S. Mikhailov, 
arXiv:cond-mat/9912054 v1 3 Dec 1999.

\bibitem{model} A. -L. Barabasi, R. Albert, H. Jeong, Physica A, {\bf 281},
69 (2000).

\bibitem{CML-critical1} S. Sinha, D. Biswas, M. Azam and S. V. Lawande,
Phys. Rev. A {\bf 46} 6242 (1992).
\bibitem{CML-critical2} V. Ahlers and A. Pikovsky, Phys. Rev. Lett. {\bf 88}
254101 (2002).

\bibitem{pre2} S. Jalan, R. E. Amritkar and C. K. Hu, unpublished.

\end{thebibliography}
\end{document}